 \documentclass[preprint2,10pt]{aastex}

\usepackage[scriptsize]{subfigure}
\usepackage{array}
\usepackage{multirow}
\usepackage{enumerate}
\usepackage{amssymb,amsmath,bm}
\shorttitle{Linearly Supporting Feature Extraction For Automated Estimation Of Stellar Atmospheric Parameters}
\shortauthors{Li et al.}

\begin{document}

\title{Linearly Supporting Feature Extraction For Automated Estimation Of Stellar Atmospheric Parameters}


%
%
%

\author{Xiangru Li\altaffilmark{1}, Yu Lu\altaffilmark{1}, Georges Comte\altaffilmark{2,3}, Ali Luo\altaffilmark{3}, Yongheng Zhao\altaffilmark{3}, Yongjun Wang\altaffilmark{1}
%
}
\affil{\altaffilmark{1}School of Mathematical Sciences, South China Normal University, 510631 Guangzhou, PR China;xiangru.li@gmail.com}
\affil{\altaffilmark{2}Aix-Marseille Universit$\acute{e}$, CNRS, Institut Pyth$\acute{e}$as, L.A.M. (Laboratoire d'Astrophysique de Marseille), UMR 7326, F-13388 Marseille Cedex, France}
\affil{\altaffilmark{3}Key Laboratory of Optical Astronomy, National Astronomical Observatories, Chinese Academy of Sciences, 100012 Beijing, PR China}




\begin{abstract}
We describe a scheme to extract linearly supporting (LSU) features from stellar spectra to automatically estimate the atmospheric parameters $T_\texttt{eff}$, log$~g$, and [Fe/H]. ``Linearly supporting'' means that the atmospheric parameters can be accurately estimated from the extracted features through a linear model. The successive steps of the process are as follow: first, decompose the spectrum using a wavelet packet (WP) and represent it by the derived decomposition coefficients; second, detect representative spectral features from the decomposition coefficients using the proposed method Least Absolute Shrinkage and Selection Operator (LARS)$_{\texttt{bs}}$; third, estimate the atmospheric parameters $T_\texttt{eff}$, log$~g$, and [Fe/H] from the detected features using a linear regression method. One prominent characteristic of this scheme is its ability to evaluate quantitatively the contribution of each detected feature to the atmospheric parameter estimate and also to trace back the physical significance of that feature. This work also shows that the usefulness of a component depends on both wavelength and frequency. The proposed scheme has been evaluated on both real spectra from the Sloan Digital Sky Survey (SDSS)/SEGUE and synthetic spectra calculated from Kurucz's NEWODF models. On real spectra, we extracted 23 features to estimate $T_\texttt{eff}$, 62 features for log$~g$, and 68 features for [Fe/H]. Test consistencies between our estimates and those provided by the Spectroscopic Sarameter Pipeline of SDSS show that the mean absolute errors (MAEs) are 0.0062 dex for log$~T_\texttt{eff}$ (83 K for $T_\texttt{eff}$), 0.2345 dex for log$~g$, and 0.1564 dex for [Fe/H]. For the synthetic spectra, the MAE test accuracies are 0.0022 dex for log$~T_\texttt{eff}$ (32 K for $T_\texttt{eff}$), 0.0337 dex for log$~g$, and 0.0268 dex for [Fe/H].

\end{abstract}


\section{INTRODUCTION}\label{Sec:Introduction}

Large-scale, deep sky survey programs, such as the Sloan Digital Sky Survey \citep[SDSS;][]{Journal:York:2000,Journal:Ahn:2012}, the Large Sky Area Multi-object Fiber Spectroscopic Telescope (LAMOST)/Guoshoujing Telescope \citep{Journal:Zhao:2006,Journal:Cui:2012}, and \emph{Gaia}-ESO Survey  \citep{Journal:Gilmore:2012,Journal:Randich:2013}, are collecting and will obtain very large numbers of stellar spectra. This enormous wealth of data makes it necessary to use a fully automated process to characterize the spectra, which in turn will enable statistical exploration of the atmospheric parameter-related properties in the spectra.

This paper investigates the problem of representing stellar spectra using a limited number of significant features to estimate atmospheric parameters. The spectrum representation problem is a vital procedure in the aforementioned tasks and is usually referred to as feature extraction in data mining and pattern recognition. For example, in atmospheric parameter estimation, a spectrum can be represented by the full observed spectrum \citep{Journal:Bailer-Jones:2000,Journal:Shkedy:2007}, the corrected spectrum \citep{Journal:Prieto:2006}, the description of some critical spectral lines \citep{Journal:Mishenina:2006,Journal:Muirhead:2012}, a statistical description \citep{Journal:Fiorentin:2007}, etc. In the present paper, we will describe a scheme for extracting LSU (linearly supporting) features from stellar spectra to estimate atmospheric parameters.

``Linearly supporting'' means that the atmospheric parameters should be accurately estimated from the extracted features using a linear model. Such a model helps to evaluate the contribution of each feature to the atmospheric parameter estimate and also to trace back the physical interpretation of that feature. It is known that there exists a high nonlinearity in the dependency of the three basic atmospheric parameters $T_\texttt{eff}$, log$~g$, and [Fe/H] on the stellar spectra \citep[Tables 6, 10 and 11 in ][]{Journal:Li:2014}. Therefore, we will first perform a nonlinear transformation on the spectrum before detecting LSU features. In this work, this initial transformation is performed using a wavelet packet (WP). The time-frequency localization of the WP allow us to isolate potential unwanted influence from noise and redundancy, and also help us to backtrack the physical absorptions or emissions that contribute to a specific analysis result. This work also shows that the effectiveness of a component depends on both wavelength and frequency.

Based on the WP decomposition of a spectrum and the Least Absolute Shrinkage and Selection Operator (LASSO) method \citep{Journal:Tibshirani:1996}, we propose an algorithm, LASSO(LARS)$_{\texttt{bs}}$, to explore a parsimonious representation of the parameterization model. Using LASSO(LARS)$_{\texttt{bs}}$, we extracted 23 features to estimate $T_\texttt{eff}$, 62 features for log$~g$, and 68 features for [Fe/H]. Experiments (Section \ref{Sec:estimation}) on real spectra from SDSS and synthetic spectra show the effectiveness of the detected features through the application of two typical linear regression methods: Ordinary Least Square (OLS) and Support Vector Regression with a linear kernel \citep[SVR$_l$;][]{Journal:Schokopf:2002,Journal:Smola:2004}.

The proposed scheme is a type of statistical learning method. The fundamental suppositions are that (1) two stars with different atmospheric parameters have distinct spectra, and (2) there is a set of observed stellar spectra or synthetic spectra with known atmospheric parameters, referred to as a training set in machine learning and data mining. Apart from the two above suppositions, there are no other a priori physical assumption. The first supposition states that there exists a mapping from stellar spectra to their atmospheric parameters. Based on these two suppositions, the proposed scheme can automatically discover this mapping, which is also known as the spectral parameterization model in astronomical data analysis, using several proposed procedures.

This paper is organized as follows. Section \ref{Sec:Data} describes the stellar spectra used in this study. In Section \ref{Sec:Estimation_LASSO}, a proposed stellar parameter estimation model is introduced. Section \ref{Sec:Spectrum_Decomposition} presents the overall configuration of the proposed scheme and investigates the feature recombination of a spectrum based on the WP transform. Section \ref{Sec:estimation} reports some experimental evaluations. Section \ref{Sec:More:Technical} discusses some technical problems, such as the optimal configuration for the WP decomposition, the sufficiency and compactness of the detected features, and the advantages and disadvantages of redundancy. Finally, we summarize our work in Section \ref{Sec:Conclusion}.

\section{DATA SETS}\label{Sec:Data}

The scheme proposed in Sections \ref{Sec:Estimation_LASSO}-\ref{Sec:More:Technical} below has been evaluated on both real spectra from SDSS/SEGUE and synthetic spectra calculated from Kurucz's NEWODF models. Real data usually present some disturbances arising from noise and pre-processing imperfections (e.g. sky lines and/or cosmic ray removal residuals, residual calibration defects), which are not present in synthetic spectra. These disturbances must be acceptable for the atmospheric parameter estimation process. Synthetic spectra are built from ground-truth parameters as reference.

Our scheme belongs to the class of statistical learning methods. The fundamental idea is to discover the linearly predictive relationship between stellar spectra and the atmospheric parameters $T_\texttt{eff}$, log$~g$, and [Fe/H] from empirical data, which constitutes a training set. At the same time, the performance of the discovered predictive relationships should also be evaluated objectively. Therefore, a separate, independent set of stellar spectra is needed for this evaluation, usually referred to as a test set in machine learning. However, most learning methods tend to overfit the empirical data. In other words, statistical learning methods can unravel some of the alleged relationships from the training data that do not hold in general. In order to avoid overfitting, we require a third independent set of spectra to optimize the parameters which need to be adjusted objectively when investigating the potential relationships: this third spectra set along with their reference parameters constitute the validation set.

Therefore, in each experiment, we will split the total spectra samples into three subsets: the training set, validation set, and test set. The training set is the carrier of knowledge and the proposed scheme should learn from this training set. The validation set is the mentor/instructor of the proposed scheme which can independently and objectively provide some advice in the learning process. The training set and validation set are used to establish a model, while the test set acts as a referee to objectively evaluate the performance of the established model. The roles of the three subsets are listed in Table \ref{Tab:DataSets:roles}.

\begin{table*}\scriptsize
\centering
\caption{Roles of the Three Data Sets. }
\begin{tabular}{ p{2cm}<{\centering} p{10cm}  }
  \hline \hline
Data Sets         &\qquad \qquad \qquad \qquad  Roles                 \\ \hline
Training Set      & To be used in
                    \begin{enumerate} \setlength{\itemsep}{0pt} \setlength{\parsep}{0pt} \setlength{\parskip}{0pt}
                    \item[(1)] Detecting features by LASSO(LARS)$_{\texttt{bs}}$;
                    \item[(2)] Estimating the parameterizing model OLS, SVR$_l$.
                    \end{enumerate}\\
Validation Set    & To be used in
 \begin{enumerate} \setlength{\itemsep}{0pt} \setlength{\parsep}{0pt} \setlength{\parskip}{0pt}
                    \item[(1)] Determining the configuration of wavelet packet decomposition (Section \ref{Sec:More:Technical:Config_wavelet_Decom});
                    \item[(2)] Determining the parameters in SVR$_l$.
                    \end{enumerate}\\                                \\
Test Set          & To be used in performance evaluation (Sections \ref{Sec:estimation} and \ref{Sec:More:Technical:Sufficiency_compactness}).                                  \\ \hline
\\
\multicolumn{2}{l}{Note. SVR$_l$: support vector regression with a linear kernel.}
\end{tabular}\label{Tab:DataSets:roles}\\
\end{table*}

\subsection{Real Spectra from SDSS/SEGUE}\label{Sec:Data:SDSS}

In this work, we use 50,000 real spectra from the SDSS/SEGUE database \citep{Journal:Abazajian:2009,Journal:Yanny:2009}. The selected spectra span the ranges [4088,9740] K in effective temperature $T_{\texttt{eff}}$, [1.015, 4.998] dex in surface gravity log$~g$, and [-3.497, 0.268] dex in metallicity [Fe/H], as given by the SDSS/SEGUE Spectroscopic Parameter Pipeline \citep[SSPP;][]{Journal:Beers:2006,Journal:Lee:2008:a,Journal:Lee:2008:b,Journal:Prieto:2008,Journal:Smolinski:2011,Journal:Lee:2011}. All stellar spectra are initially shifted to their rest frames (zero radial velocity) using the radial velocity provided by SSPP. They are also rebinned to a maximal common log(wavelength) range [3.581862, 3.963961] with a sampling step of 0.0001.\footnote{The common wavelength range is approximately [3818.23, 9203.67]${\AA}$.} The sizes of the training set, validation set, and test set are 10,000, 10,000 and 30,000 spectra, respectively.

We take the real spectra atmospheric parameters previously estimated by SSPP as reference values. The SSPP estimation is based on both stellar spectra and \emph{ugriz} photometry by combining the results of multiple techniques to alleviate the limitations of a specific method, see \citet{Journal:Lee:2008:a} and references therein. SSPP has been extensively validated by comparing its estimates with the sets of parameters obtained from high-resolution spectra from SDSS-I/SEGUE stars \citep{Journal:Prieto:2008} and with the available information from the literature for stars in Galactic open and globular clusters \citep{Journal:Lee:2008:b,Journal:Smolinski:2011}.

\subsection{Synthetic Spectra}\label{Sec:Data:Synthetic}

A set of 18,969 synthetic spectra are calculated from the SPECTRUM (v2.76) package \citep{Con:Gray:1994} with Kurucz's NEWODF models \citep{Journal:Castelli:2003}. When generating the synthetic spectra, 830,828 atomic and molecular lines are used (contained in two files luke.lst and luke.nir.lst); the atomic and molecular data are stored in the file stdatom.dat, which includes solar atomic abundances from \citet{Journal:Grevesse:1998}. The SPECTRUM package and the three data files can be downloaded from website.\footnote{http://stellar.phys.appstate.edu/spectrum/download.html.}

Our grids of synthetic stellar spectra span the parameter ranges [4000,9750] K in $T_\texttt{eff}$ (45 values, step sizes of 100K between 4000 and 7500 and 250 K between 7750 and 9750K), [1, 5] dex in log$~g$ (17 values, step size of 0.25 dex), and [-3.6, 0.3] dex in [Fe/H] (27 values, step size of 0.2 dex between -3.6 and -1 dex 0.1 dex between -1 and 0.3 dex). The synthetic stellar spectra are also split into three subsets: the training set, validation set, and test set with respective sizes of 8500, 1969 and 8500 spectra.

\section{A LINEAR ESTIMATION MODEL FOR ATMOSPHERIC PARAMETERS}\label{Sec:Estimation_LASSO}

\subsection{Model}\label{Sec:Estimation_LASSO:Model}
Let a vector $\bm{x} = (x_{1}, \cdots, x_{p})^T$ represent a spectrum and $y$ be an atmospheric parameter to be estimated, where $p > 0$. The component $x_j$ represents the flux of the spectrum $\bm{x}$, $j \in \{1, 2, \cdots, p\}$. We investigate the atmospheric parameter estimation problem based on a linear model:
\begin{equation}\label{Equ:linear:model}
   \hat{y} = f(\bm{x};\bm{w}) = \sum_{j=1}^{p}{w_j x_j},
\end{equation}
where $\bm{w}=(w_1,\cdots,w_{p})$ are free parameters characterizing the model. For convenience, we assume that the mean of the variable $y$ to be estimated is zero, otherwise a $w_0$ should be added to the right side of Equation (\ref{Equ:linear:model}).

In this work, $y$ can be the effective temperature $T_\texttt{eff}$, the surface gravity log$~g$, or the metallicity [Fe/H]. The stellar spectra are analyzed three times, respectively, for these three parameters. To reduce the dynamical range and to better represent the uncertainties of the spectral data, we use log $T_{\texttt{eff}}$ instead of $T_{\texttt{eff}}$ in our analysis \citep{Journal:Fiorentin:2007}.

Under the linear regression model in Equation (\ref{Equ:linear:model}), it is easy to evaluate the influence from a flux component $x_j$ on the estimate $\hat{y}$: a regression coefficient $w_j$ provides the variation of the parameter $y$ to be estimated when the component $x_j$ is changed by one unit while the other flux components $\{x_{1}, \cdots, x_{j-1}, x_{j+1}, \cdots, x_{p}\}$ are kept constant. Therefore, the model in Equation (\ref{Equ:linear:model}) describes the linear support for the parameter to be estimated from every component of a spectrum.

Suppose that $S_{F}$ is a set consisting of the flux/predictor components of stellar spectra whose model coefficients $w_j \neq 0$ in Equation (\ref{Equ:linear:model}), and $\overline{S}_{F}$ is a set consisting of the flux components whose model coefficients $w_j = 0$. Then, all of the components belonging to $\overline{S}_{F}$ are ineffective in model (\ref{Equ:linear:model}), and $S_{F}$ is the set of components necessary and sufficient for estimating $y$ based on the linear  model (\ref{Equ:linear:model}). Therefore, the components in $S_F$ are called a set of LSU features for the parameter to be estimated in Equation (\ref{Equ:linear:model}).

\subsection{Model Selection}\label{Sec:sub:Model_selection}
The model in Equation (\ref{Equ:linear:model}) can be determined by checking its consistency with a set of labeled spectra
\begin{equation}\label{Equ:training_sp}
S = \{ (\bm{x}^i, y_i), i = 1, 2, \cdots, N \},
\end{equation}
where $\bm{x}^i$ is a spectrum and $y_i$ is an atmospheric parameter. The consistency is usually evaluated using the Mean of Squared Error (MSE):
\begin{equation}\label{Equ:MSE}
  \begin{split}
  MSE(\bm{w}) = &\frac{1}{N} \sum_{i=1}^{N}{(y_i -\hat{y}_i)^2} \\
         = &\frac{1}{N} \sum_{i=1}^{N}{(y_i -f(\bm{x}^i; \bm{w}))^2}.
  \end{split}
\end{equation}

When we select the model in Equation (\ref{Equ:linear:model}) by minimizing the MSE error
\begin{equation}\label{Sec:MSE:Obj}
   \hat{\bm{w}} = arg\min\limits_{\bm{w}}{\{MSE(\bm{w})\}},
\end{equation}
the model $f(\cdot;\hat{\bm{w}})$ derived by Equation (\ref{Sec:MSE:Obj}) is referred to as the OLS regression. In this OLS model, most of the coefficients $\hat{w}_1, \hat{w}_2, ...\hat{w}_n$ are non-zero and we will call it a complex model for convenience. This complexity usually leads model (\ref{Equ:linear:model}) to suffer from redundancy and irrelevant variables in the data (as noise or pre-processing artefacts), which in turn can lead to overfitting and difficulties in exploring the most significant factors in high-dimensional spectra.

To overcome or alleviate the aforementioned limitations, a typical strategy is to regularize the object function (\ref{Sec:MSE:Obj}) by the $\ell_{1}$-norm of the model parameter $w$
\begin{equation}\label{Equ:MSE:LASSO}
   \hat{\bm{w}} = arg\min\limits_{\bm{w}}{\{  \sum_{i=1}^{N}{(y_i -f(\bm{x}^i; \bm{w}))^2} +  \lambda \|\bm{w}\|_1\}},
\end{equation}
where
\begin{equation}
    \|\bm{w}\|_1  = \sum_{i=1}^{p}{|w_{i}|}.
\end{equation}
The model $f(\cdot;\hat{\bm{w}})$ derived from Equation (\ref{Equ:MSE:LASSO}) is called LASSO (Least Absolute Shrinkage and Selection Operator) \citep{Journal:Tibshirani:1996}. Here, $\lambda \geq 0$ is a tuning parameter that controls the amount of non-zero parameters $w_i$, or equivalently the complexity of the selected model. Studies show that LASSO can effectively filter out most of the redundant or irrelevant variables by shrinking some parameters $w_i$ to exactly zero \citep{Book:James:2013}. We use the Matlab implementation \citep{Software:Sjostrand:2005} of LASSO based on the LARS algorithm \citep{Journal:Efron:2004}. To highlight the implementation based on LARS, we label it LASSO(LARS). In LASSO(LARS), the parameter $\lambda$ can be equivalently replaced with the number $m$ of non-zero parameters $w_i$ \citep{Journal:Efron:2004}.

Selecting features using LASSO is equivalent to determining the subset of model coefficients $\{w_j, j= 1, \cdots, p\}$ with non-zero values. Suppose that $S_W$ represents the subset of model coefficients $w_j$ in equations (\ref{Equ:linear:model}) and (\ref{Sec:MSE:Obj}). Essentially, LARS is an implementation of LASSO based on a forward selection scheme. It starts with all coefficients equal to zero $\{ w_j = 0, j= 1, \cdots, p \}$ by setting $\hat{y}_0 =0$ \footnote{The subscript $0$ of $\hat{y}_0$ indicates that this estimate is computed without considering any predictor.} and $S_W =\varnothing$, where $\varnothing$ is the empty set. It then expands $S_W$ gradually as follows. First, the LARS algorithm tries to find the predictor $x_{j_1}$ best correlated with the response $y$ and expands $S_W$ from the empty set to $\{w_{j_1}\}$ by setting the value of $w_{j_1}$ to move by the largest possible step in the direction of predictor $x_{j_1}$ until some other predictor $x_{j_2}$ has as much correlation with the current estimation residual\footnote{Now, $x_{j_1}$ and $x_{j_2}$ are tied for the highest correlation with the current estimate residual.}. In the next step, LARS stops the motion along $x_{j_1}$, proceeds in an equiangular direction between the two predictors $x_{j_1}$ and $x_{j_2}$ (least angle direction) by adjusting $w_{j_1}$ and $w_{j_2}$ simultaneously until a third predictor $x_{j_3}$ has as much correlation with the current estimate residual\footnote{Currently, $x_{j_1}$, $x_{j_2}$, and $x_{j_3}$ are tied for the highest correlation with the current estimate residual.}, and setting $S_W = \{w_{j_1}, w_{j_2}\}$. Then, LARS proceeds in an equiangular direction between $x_{j_1}$, $x_{j_2}$ and $x_{j_3}$ (least angle direction) until a fourth predictor $x_{j_4}$ is found, and $S_W = \{w_{j_1}, w_{j_2}, w_{j_3}\}$. The LARS algorithm can select $m$ features if the above procedure continues, where $m$ is an empirically preset number representing the number of non-zero parameters $w_i$. Interested readers are referred to \citet{Journal:Efron:2004} for further information concerning LARS.

Note that aside from LASSO, there are multiple alternatives for sparse model selection, for example, Forward Stepwise Selection, Backward Stepwise Selection, Forward Stagewise \citep{Book:Hastie:2009,Book:James:2013}, Elastic Net \citep{Journal:Zou:2005}, etc.

\subsection{Refining the Selected Model}\label{Sec:sub:Model_Further}
To select a model with $k_0$ features, we can use LASSO(LARS) directly to impose the constraint $k_0$ on the features number, or first select a model with $m$ features using LASSO(LARS), and then eliminating the $m - k_0$ features iteratively one by one, where $k_0$ and $m$ are two positive integers and $m \ge k_0$. For convenience, we call the above-mentioned schemes, respectively, ``direct LASSO(LARS)'' and ``LASSO(LARS)$_{\texttt{bs}}$'' (LASSO(LARS) with backward selection). Experiments show that the LASSO(LARS)$_{\texttt{bs}}$ scheme is better than the direct LASSO(LARS).

On the whole, LASSO(LARS) is a forward selection method. Its drawback is that each addition of a new feature may make one or more of the already included variables not sufficiently significant, and even less significant than the excluded variables. The LASSO(LARS)$_{\texttt{bs}}$ can choose more variables as candidates and take more combination effects of variables into consideration. This is a balance between accuracy and time complexity. The proposed LASSO(LARS)$_{\texttt{bs}}$ scheme works as follows.
\begin{enumerate}[1.]
\item Select a linear model with $m$ non-zero coefficients based on a training set by LASSO(LARS) (see Equation (\ref{Equ:MSE:LASSO}) and Section \ref{Sec:sub:Model_selection} above); the corresponding variables form a set $S_F$.
\item For every element $s \in S_F$, compute two OLS estimates $f(\cdot;\hat{\bm{w}}_1^s)$ and $f(\cdot;\hat{\bm{w}}_2^s)$ based on the variables $S_F$ and $S_F - \{s\}$, respectively, from a training set.
\item Evaluate the effectiveness of $s$ using $\texttt{Eff}(s) = MAE(f(\cdot;\hat{\bm{w}}_2^s)) - MAE(f(\cdot;\hat{\bm{w}}_1^s))$, where the MAE is computed based on a validation set (see below Equation (\ref{Equ:MAE}) in Section \ref{Sec:sub:Evaluation_methods} for the definition of the MAE).
\item Compute $s_0 = arg\min\limits_{s\in S_F}${\{\texttt{Eff}(s)\}}, and let $S_F = S_F - \{s_0\}$.
\item If the size of $S_F$ is greater than $k_0$, go to step 2; otherwise, return $S_F$ as the extracted features and take the OLS estimate of $S_F$ as the final model.
\end{enumerate}

\subsection{Evaluation Methods}\label{Sec:sub:Evaluation_methods}
Suppose that $S_{te} = \{ (\bm{x}^m, y_m), m = 1, 2, \cdots, M \}$ is a test set. In this work, the performance of the proposed scheme is evaluated using three methods: Mean Error (ME), MAE, and Standard Deviation (SD). They have been used in related research \citep{Journal:Fiorentin:2007,Journal:Jofre:2010,Journal:Tan:2013} and are defined as follows:
\begin{equation}\label{Equ:ME}
   ME = \frac{1}{M}\sum_{m=1}^{M}{e_{m}},
\end{equation}
\begin{equation}\label{Equ:MAE}
   MAE = \frac{1}{M}\sum_{m=1}^{M}|e_{m}|,
\end{equation}
\begin{equation}\label{Equ:SD}
   SD = \sqrt{\frac{1}{M}\sum_{m=1}^{M}(e_{m}-ME)^2},
\end{equation}
where $e_{m}$ is the error/difference between the reference value of the stellar parameter and its estimate
\begin{equation}\label{Equ:deviation}
   e_m = y_m - f(\bm{x}^m), ~m = 1,~\cdots, M.
\end{equation}

ME, MAE, and SD are all widely used in the performance evaluation of an estimation process. Each evaluation method focuses on different aspects of the estimation process. ME measures the average magnitude of the deviation, reflecting systematic errors: if the expectation of ME is 0, then $f(\bm{x}^m)$ is referred to as a statistically unbiased estimator of $y_m$. MAE accesses the average magnitude of the deviation by ignoring the sign/direction of an error. SD shows how much variation exists in an estimation error and reflects the stability/robustness of the estimation process. A low SD indicates that the performance of the proposed estimation scheme is very stable; a high SD indicates that its performance is sensitive to a specific spectrum to be processed. If the errors $\{e_m, m = 1, 2, \cdots, M\}$ are independent, identically distributed (iid) random variables with  a normal density distribution $\phi(e;\mu,\sigma)= \frac{1}{\sigma \sqrt{2\pi}}e^{-\frac{(e-\mu)^2}{2\sigma^2}}$, then ME and SD are the estimates of $\mu$ and $\sigma$, respectively. In addition, MAE is the estimation of $\sqrt{\frac{2}{\pi}} \sigma$ if the errors $\{e_m, m = 1, 2, \cdots, M\}$ are iid random variables with a normal density distribution $\phi(e;0,\sigma)= \frac{1}{\sigma \sqrt{2\pi}}e^{-\frac{e^2}{2\sigma^2}}$ \citep{Journal:Geary:1935}.

\section{OVERALL CONFIGURATION AND SPECTRAL FEATURE ANALYSIS BASED ON WP}\label{Sec:Spectrum_Decomposition}
\subsection{Overall Configuration}\label{Sec:Spectrum_Decomposition:Overall_Config}

There exists a high nonlinearity in the dependence of the atmospheric parameters $T_\texttt{eff}$, log$~g$, [Fe/H] on stellar spectra \citep[Table 6, Table 10, Table 11 in ][]{Journal:Li:2014}. Therefore, a nonlinear transformation should be performed on spectra before detecting LSU features to estimate stellar parameters. Several statistical procedures will be performed to estimate the atmospheric parameters $T_\texttt{eff}$, log~$g$, and [Fe/H] in the proposed scheme.

\begin{figure*}
\begin{center}
\includegraphics[ width =3in]{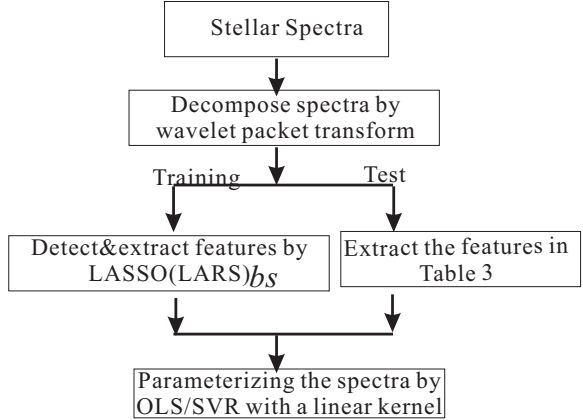}
\end{center}
\setlength{\abovecaptionskip}{1pt} \caption{Flowchart to show the order in which the statistical procedures are used in the analysis.}
\label{Fig:Flowchart}
\end{figure*}

A flowchart of the procedures is presented in Fig. \ref{Fig:Flowchart} to demonstrate the end-to-end flow in the analysis. The initial step ``Decompose spectra by WP transform'' is introduced in Section \ref{Sec:Spectrum_Decomposition:wavelet_packet_transform} below. This step requires that some technical choices be made, such as the selection of the wavelet basis function and of the level of wavelet packet decomposition (WPD). These problems are discussed in Section \ref{Sec:More:Technical:Config_wavelet_Decom}. After decomposing the stellar spectra, we can detect and extract features using the LASSO(LARS)$_{\texttt{bs}}$ method (Section \ref{Sec:sub:Model_Further}) to reduce redundancy and noise (Section \ref{Sec:Spectrum_Decomposition:feature_selection}).

\subsection{WP Transform}\label{Sec:Spectrum_Decomposition:wavelet_packet_transform}
We apply the WP transform to our stellar spectrum and decompose it into a series of components with different wavelengths and different frequencies (time-frequency localization).

Suppose that $\bm{x}= (x_1, x_2, \cdots, x_n)^T \in R^n$ is a spectrum consisting of $n$ fluxes (sampling points): we refer to it as a signal with length $n$. Since the spectrum considered is a one-dimensional signal, our discussion focuses on one-dimensional WPs.

\subsubsection{Principles}\label{Sec:Spectrum_Decomposition:wavelet_packet_transform:principles}
\begin{figure*}
\begin{center}
\includegraphics[ width =3in]{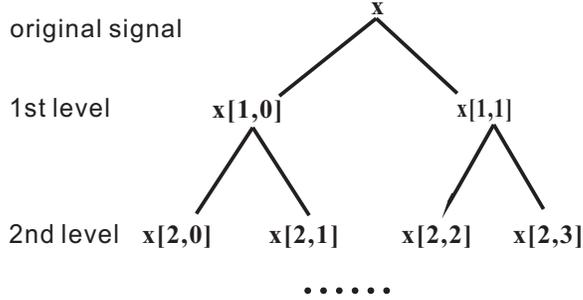}
\end{center}
\setlength{\abovecaptionskip}{1pt} \caption{WP decomposition tree: principles of WP. A signal can be decomposed into a low-frequency approximation signal and high-frequency details, and can be iteratively re-decomposed to provide increasingly accurate frequency resolution.}
\label{Fig:WP_tree}
\end{figure*}

WPs can decompose a signal into a low-frequency approximation signal and high-frequency details, and can iteratively re-decompose those signals to provide increasingly accurate frequency resolution. For example, in Fig. \ref{Fig:WP_tree}, WPs decompose a signal $\bm{x}$ into a low-frequency approximation signal $\bm{x}[1,0] = (x_{1,0}^1, x_{1,0}^2, \cdots, x_{1,0}^{n_1})$ and a high-frequency detail signal $\bm{x}[1,1] = (x_{1,1}^1, x_{1,1}^2, \cdots, x_{1,1}^{n_1})$. We call $\bm{x}[1] = \{ \bm{x}[1,0], \bm{x}[1,1] \}$ the first-level WP decomposition of signal $\bm{x}$. Then, $\bm{x}[1,0]$ can also be further decomposed into $\bm{x}[2,0] = (x_{2,0}^1, x_{2,0}^2, \cdots, x_{2,0}^{n_2})$ and $\bm{x}[2,1] = (x_{2,1}^1, x_{2,1}^2, \cdots, x_{2,1}^{n_2})$, and $\bm{x}[1,1]$ into $\bm{x}[2,2] = (x_{2,2}^1, x_{2,2}^2, \cdots, x_{2,2}^{n_2})$ and $\bm{x}[2,3] = (x_{2,3}^1, x_{2,3}^2, \cdots, x_{2,3}^{n_2})$. These four new resulting signals are called the second-level WP decomposition $\bm{x}[2] = \{ \bm{x}[2,j], j = 0, \cdots, 3  \}$ of signal $\bm{x}$.

 If this decomposition procedure is repeated again and again, then a series of decompositions, $\bm{x}[3], \bm{x}[4], \cdots$, are generated and form the WP decomposition tree of the signal $\bm{x}$ (see Fig. \ref{Fig:WP_tree}), where $\bm{x}[i] = \{ \bm{x}[i, j], j = 0, \cdots, 2^i -1 \} \in R^{N_i}$ is the $i$th level WP decomposition, where $N_i$ is an integer and is described in detail in Section \ref{Sec:Spectrum_Decomposition:wavelet_packet_transform:Implementation}. At each level,
\begin{equation}
\bm{x}[i, j] = \{x_{i,j}^k, 0\leq k\leq n_i\}
\end{equation}
is a set of decomposition components belonging to a frequency sub-band, where $n_i$ is an integer and is described in detail in Section \ref{Sec:Spectrum_Decomposition:wavelet_packet_transform:Implementation}. The frequency of a sub-band $\bm{x}[i, j_1]$ is higher than that of a sub-band $\bm{x}[i, j_2]$, where $i \geq 1$ and $0\leq j_2 < j_1 < 2^i$. Therefore, there are $2^i$ frequency sub-bands on the $i$th level WP decomposition, and the $(i+1)$th level WP decomposition has higher frequency resolution than the $i$th level WP decomposition, where $i>0, j>0$.

Traditionally, a sub-band $\bm{x}[i,j]$ is referred to as a node of a WPD tree (Fig. \ref{Fig:WP_tree}), and the component $x_{i,j}^k$ is referred to as a WP coefficient.

\subsubsection{Implementations}\label{Sec:Spectrum_Decomposition:wavelet_packet_transform:Implementation}
In this work, we use the WP implementation of Wavelet Toolbox in Matlab. WP decomposition is implemented by filtering and downsampling, and the filter is a vector associated with a basis function. Suppose that $\bm{x}$ is a signal with length $n$ to be decomposed by WP in Fig. \ref{Fig:WP_tree} and the length of the filter is $m$. Then, the length of $\bm{x}[1,j]$ is $n_1 = ceil(n/2) + ceil(m/2) - 1$, where $j \in \{0, 1\}$ and $ceil(z)$ is a function that rounds up its parameter $z$ to the nearest integer toward infinity:
\begin{equation}
   ceil(z) = k, if~~ k-1 < z \leq k,
\end{equation}
where $k$ is an integer. Therefore, the length of the first-level WP decomposition $\bm{x}[1]$ is $N_1 = n_1 \times 2^1 = 2(ceil(n/2) + ceil(m/2) - 1)$.

Similarly, if the length of a sub-band $\bm{x}[i,j]$ is $n_i$, then the length of the $i$th level WP decomposition $\bm{x}[i]$ is $N_i = n_i \times 2^i$, where $i \geq 1$, $0 \leq j < 2^i$, and $2^i$ is the number of sub-bands with different frequency at the WP decomposition level $i$; The length of a sub-band $\bm{x}[i+1, j]$ on the $(i+1)$th WP decomposition is $n_{i+1} = ceil(n_i/2) + ceil(m/2) - 1$, and the length of the $(i+1)$th level WP decomposition $\bm{x}[i+1]$ is
\begin{equation}\label{Equ:length}
   N_{i+1} = n_{i+1} \times 2^{i+1}.
\end{equation}

We investigate the feature analysis problem of WP-decomposed stellar spectra using the following typical basis functions: Biorthogonal basis (bior), Coiflets(coif), Daubechies basis (db), Haar (haar), ReverseBior (rbio), and Symlets (sym) \citep{Journal:Mallat:1989,Book:Mallat:2009,Book:Daubechies:1992}. The filters associated with these functions are, respectively, referred to as filter(bior2.2), filter(coif4), filter(db4), filter(haar), filter(rbio4.4), and filter(sym4)\citep[see Documentation of Matlab--wavelet filters: as
][]{Book:MathWorks:2014:wfilters}\footnote{There are multiple variants for basis functions bior, coif, db, rbio, and sym in the implementation of the Matlab wavelet toolbox. The numbers behind them are the indexes of the variants.}. The respective filter lengths are 6, 24, 8, 2, 10, and 8. The length of all our spectra is $n=3821$. Based on Equation (\ref{Equ:length}), the lengths of the WP decomposition $\{\bm{x}[i], i=1, \cdots, 6\}$ are presented in Table \ref{Tab:wb_filters:length} for the above-mentioned basis functions.

\begin{table*}\scriptsize
\setlength{\abovecaptionskip}{-100pt}
\setlength{\belowcaptionskip}{-100pt}
\centering
\caption{Length of wavelet packet decomposition of a spectrum used in this work based on six typical wavelet bases.}
\begin{tabular}{p{1cm}<{\centering}p{1.3cm}<{\centering}p{1.3cm}<{\centering}p{1.3cm}<{\centering}p{1.3cm}<{\centering}p{1.3cm}<{\centering}p{1.3cm}<{\centering}}
  \hline \hline
 LWPD    & filter(bior2.2)& filter(coif4)& filter(db4)&  filter(haar)  & filter(rbio4.4)& filter(sym4)\\
 \hline
  1      &    3826      &     3844     &   3828      &   3822          &     3830      &    3828      \\
  2      &    3836      &     3888     &   3840      &   3824          &     3848      &    3840      \\
  3      &    3856      &     3976     &   3864      &   3824          &     3880      &    3864      \\
  4      &    3888      &     4160     &   3920      &   3824          &     3952      &    3920      \\
  5      &    3968      &     4512     &   4032      &   3840          &     4096      &    4032      \\
  6      &    4096      &     5248     &   4224      &   3840          &     4352      &    4224      \\
\hline
\\
\multicolumn{7}{l}{Note. LWPD: level of wavelet packet decomposition.}
\end{tabular}\label{Tab:wb_filters:length}
\end{table*}

\subsubsection{Reconstruction and Visualization}\label{Sec:Spectrum_Decomposition:Rec_Vis}
As for a level $i$ in the WP decomposition tree (Fig. \ref{Fig:WP_tree}), WPD is a mapping $wpdec: R^n \rightarrow R^{N_i}$ from the spectral space $R^n$ to a WPD space $R^{N_i}$
\begin{equation}
 wpdec(\bm{x}, i) = \bm{x}[i],
\end{equation}
where $\bm{x} \in R^n$ is a spectrum and $\bm{x}[i] \in R^{N_i}$. Based on the theory of WP \citep{Book:Daubechies:1992,Book:Mallat:2009}, we can also reconstruct the spectrum $\bm{x}$ from WP decomposition $\bm{x}[i]$ by a mapping $wprec: R^{N_i} \rightarrow R^n$ \citep[see Documentation of Matlab WP reconstruction: as ][]{Book:MathWorks:2014:wprec}, this process is referred to as WP reconstruction.

\begin{figure*}
  \centering
\subfigure[A spectrum $\bm{x}$.]{
    \label{Fig:wp_dec:x} 
    \includegraphics[width =3.2in]{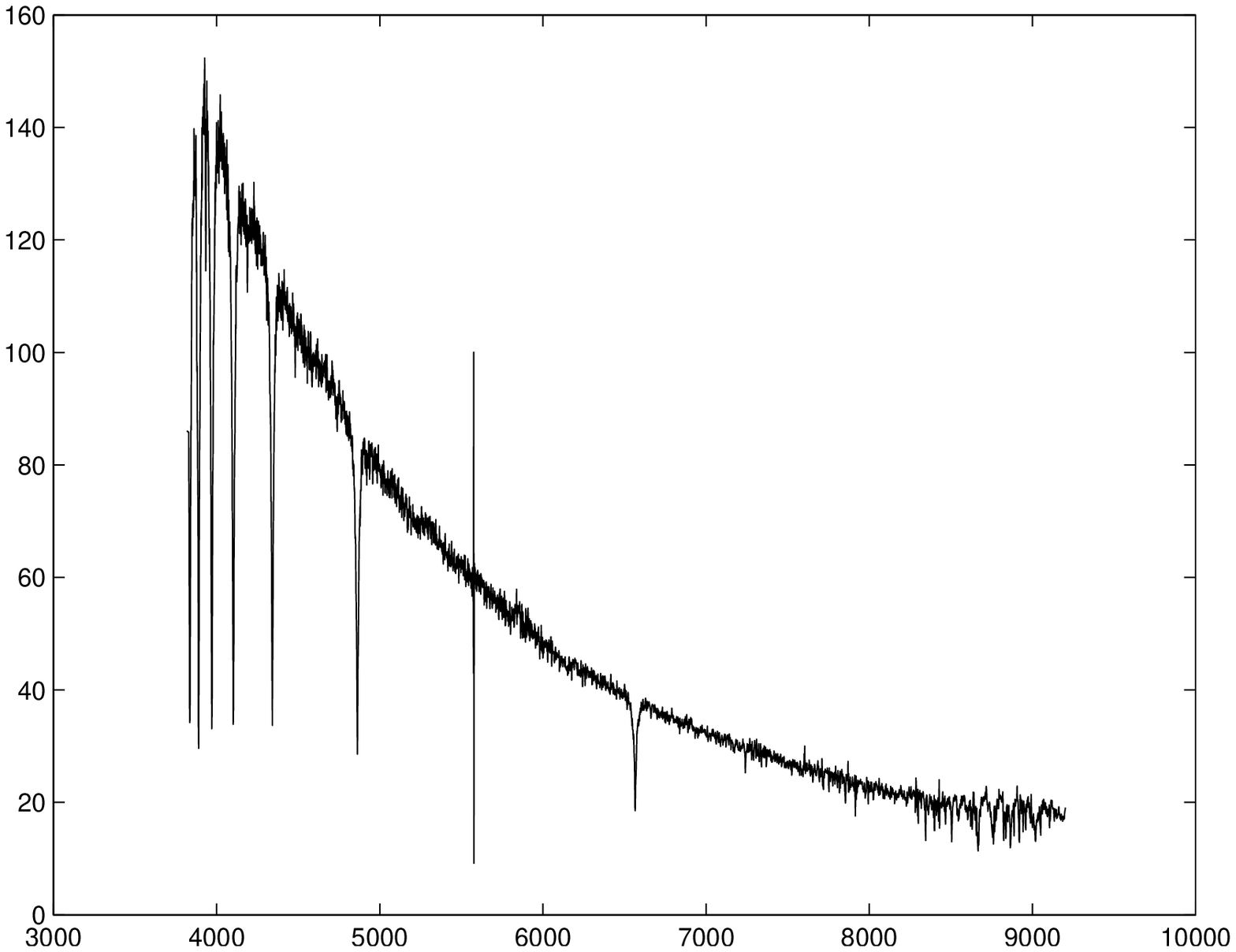}}
\hspace{-0.17in}
  \subfigure[WPD $\bm{x}{[}1,0{]}$.]{
    \label{Fig:wp_dec:x_1_0} 
    \includegraphics[width =3.2in]{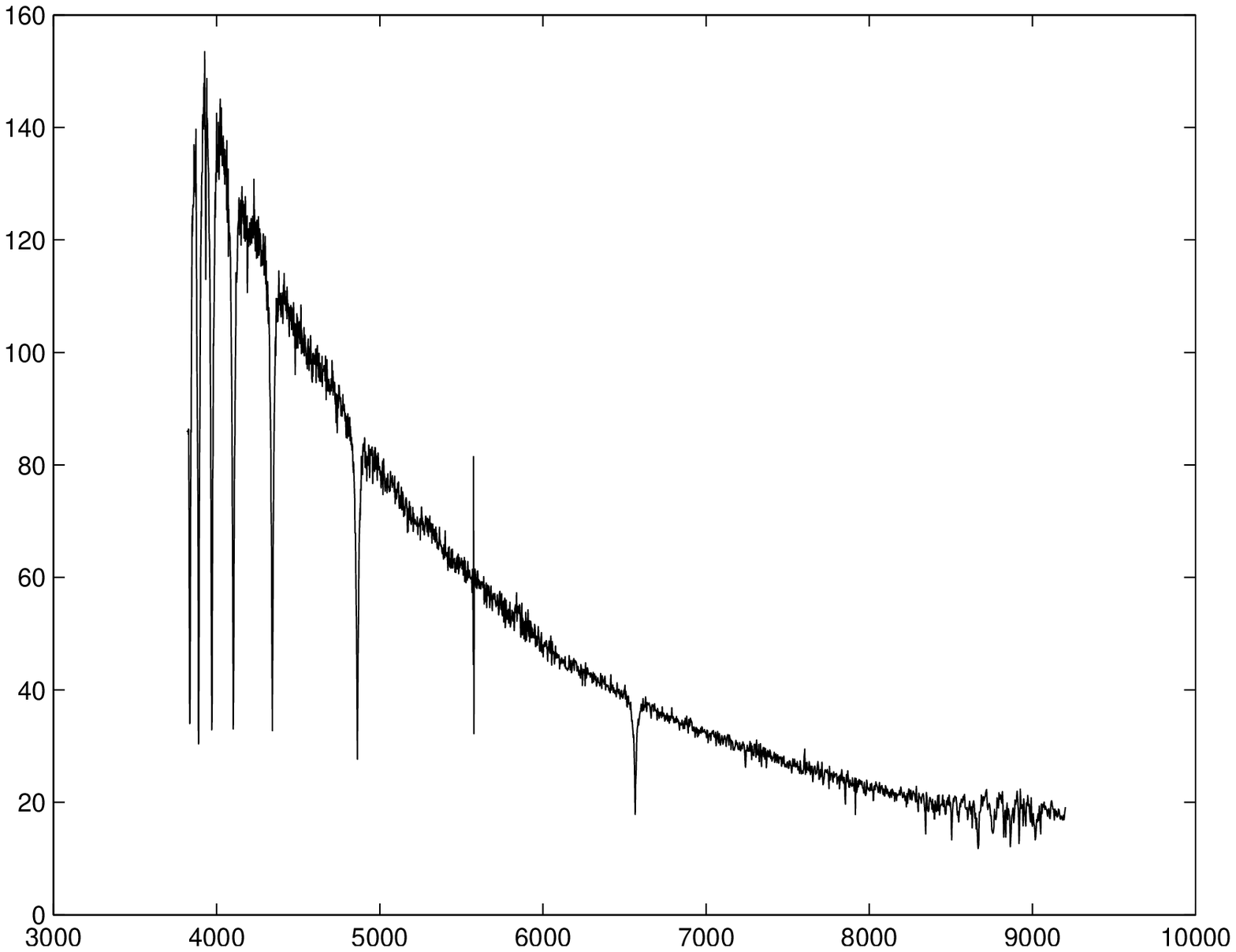}}
\hspace{-0.17in}
  \subfigure[WPD  $\bm{x}{[}1,1{]}$.]{
    \label{Fig:wp_dec:x_1_1} 
    \includegraphics[width =3.2in]{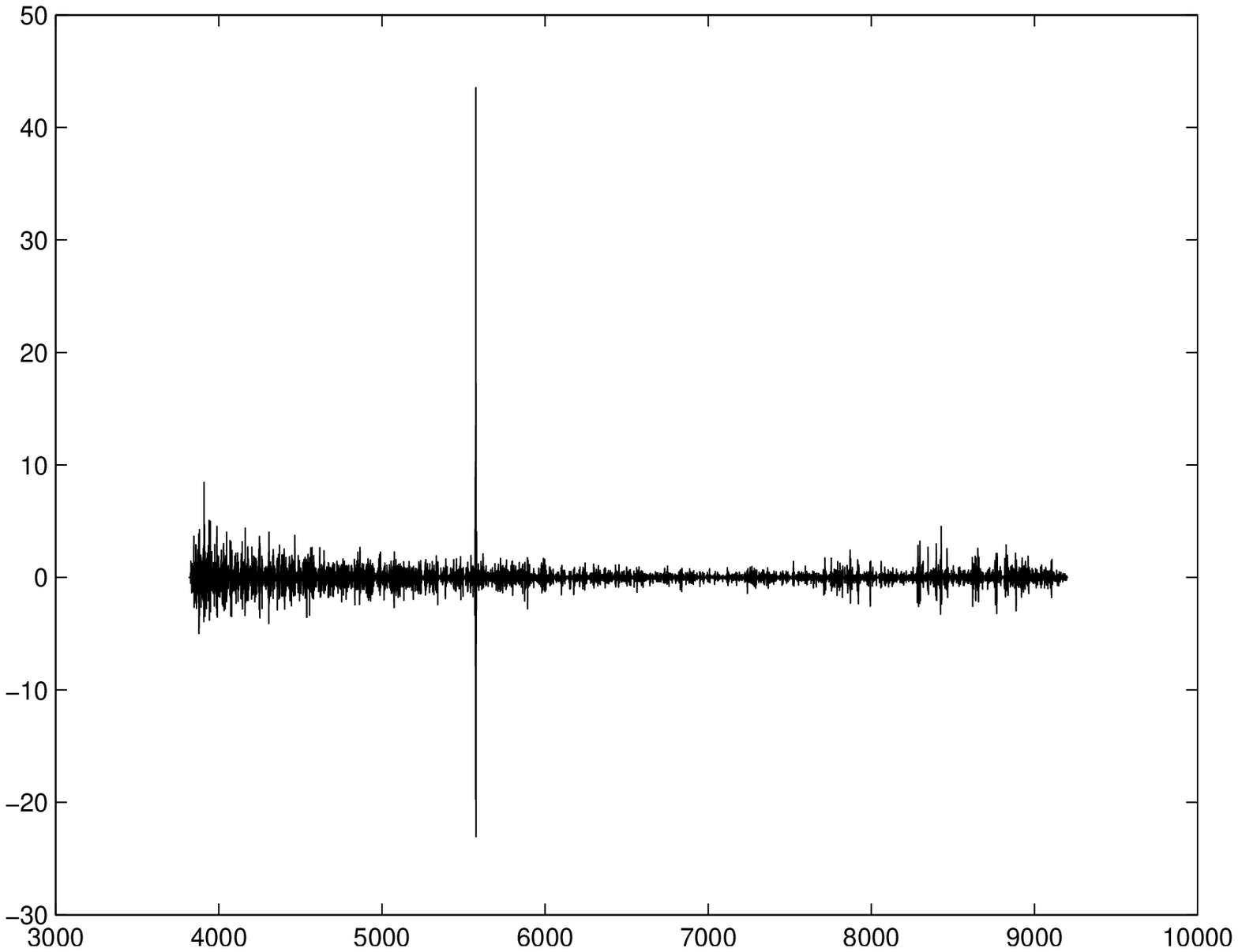}}
\hspace{-0.17in}
  \subfigure[WPD  $\bm{x}{[}2,0{]}$.]{
    \label{Fig:wp_dec:x_2_0} 
    \includegraphics[width =3.2in]{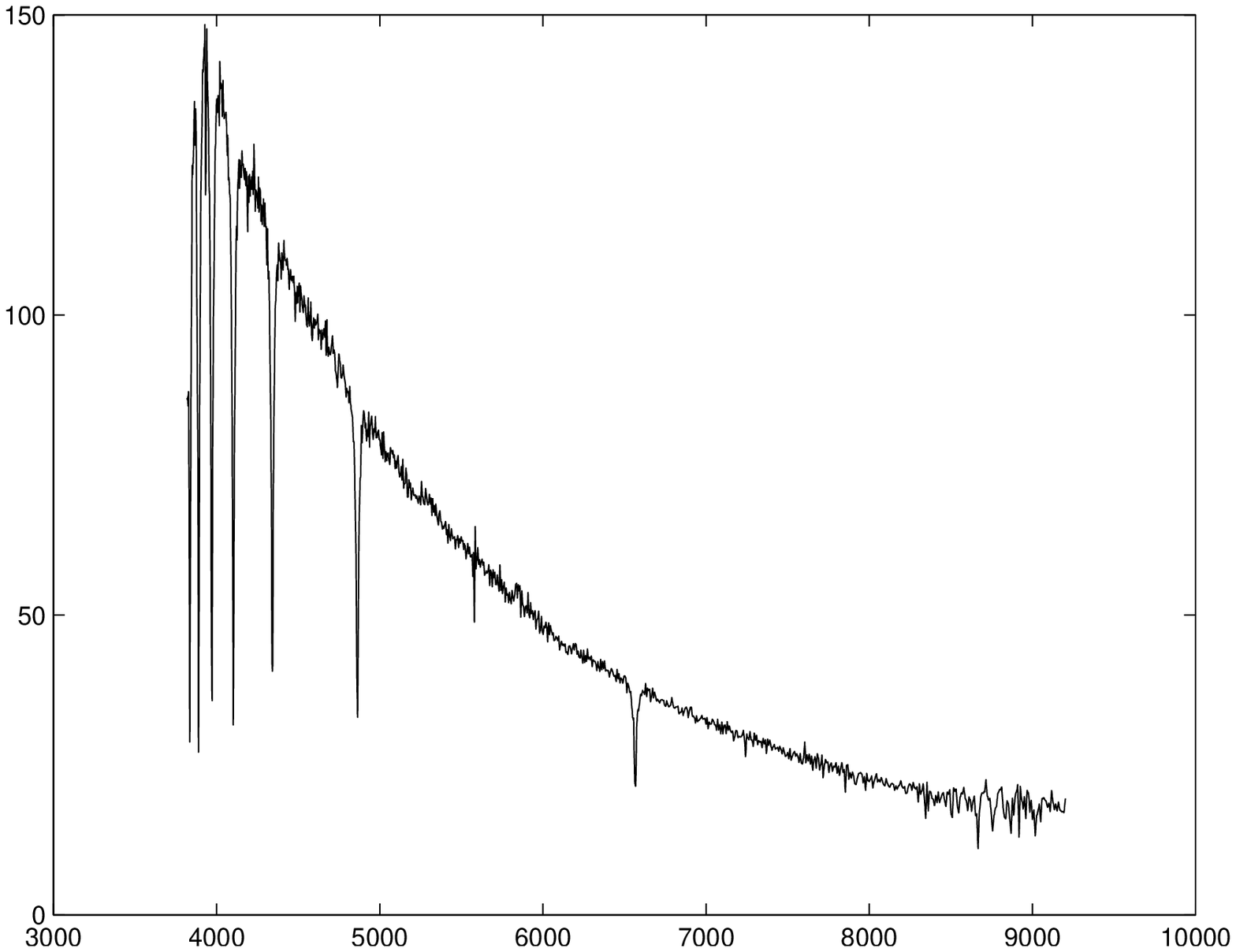}}
\hspace{-0.17in}
  \subfigure[WPD  $\bm{x}{[}2,1{]}$.]{
    \label{Fig:wp_dec:x_2_1} 
    \includegraphics[width =2.1in]{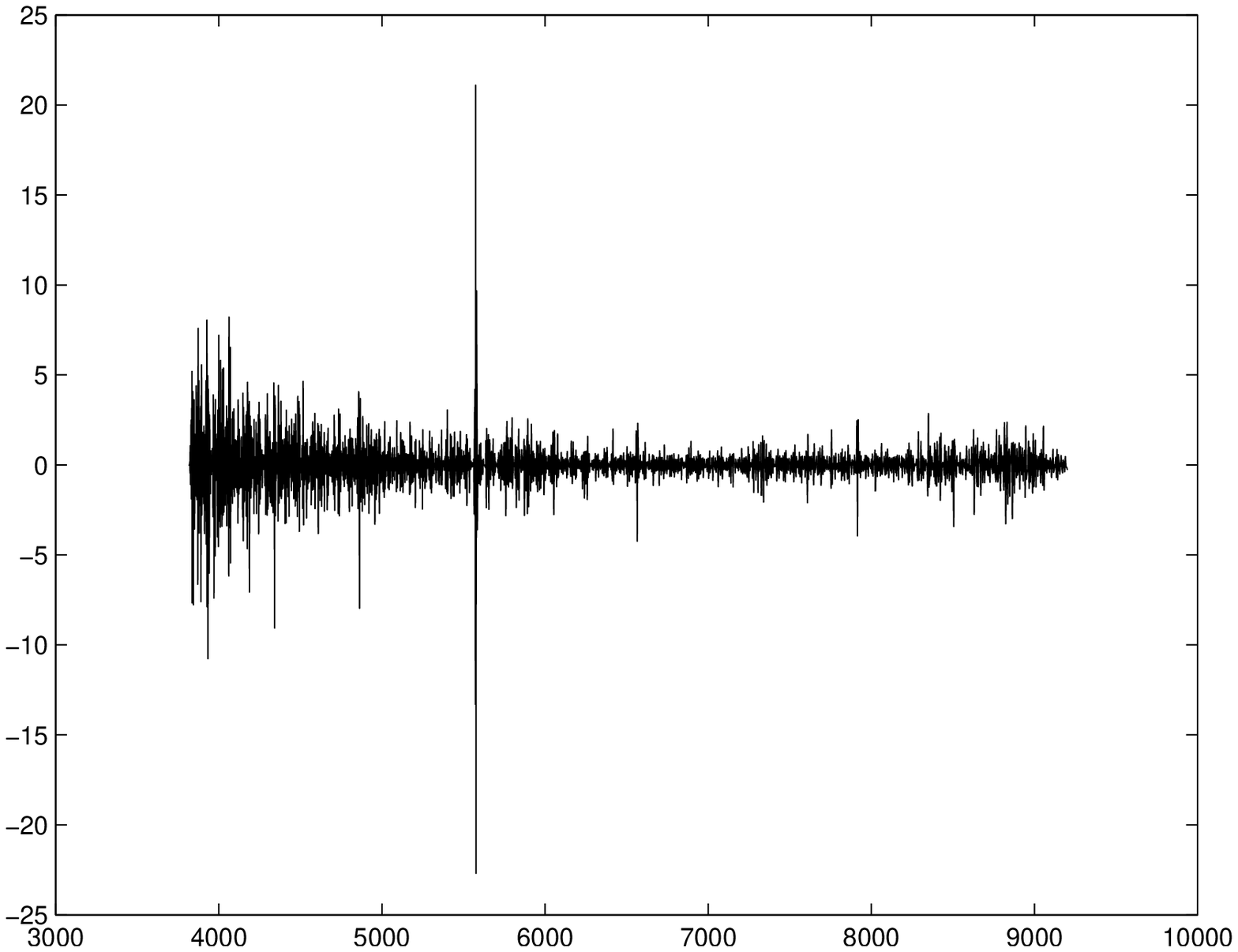}}
\hspace{-0.17in}
    \subfigure[WPD  $\bm{x}{[}2,2{]}$.]{
    \label{Fig:wp_dec:x_2_2} 
    \includegraphics[width =2.1in]{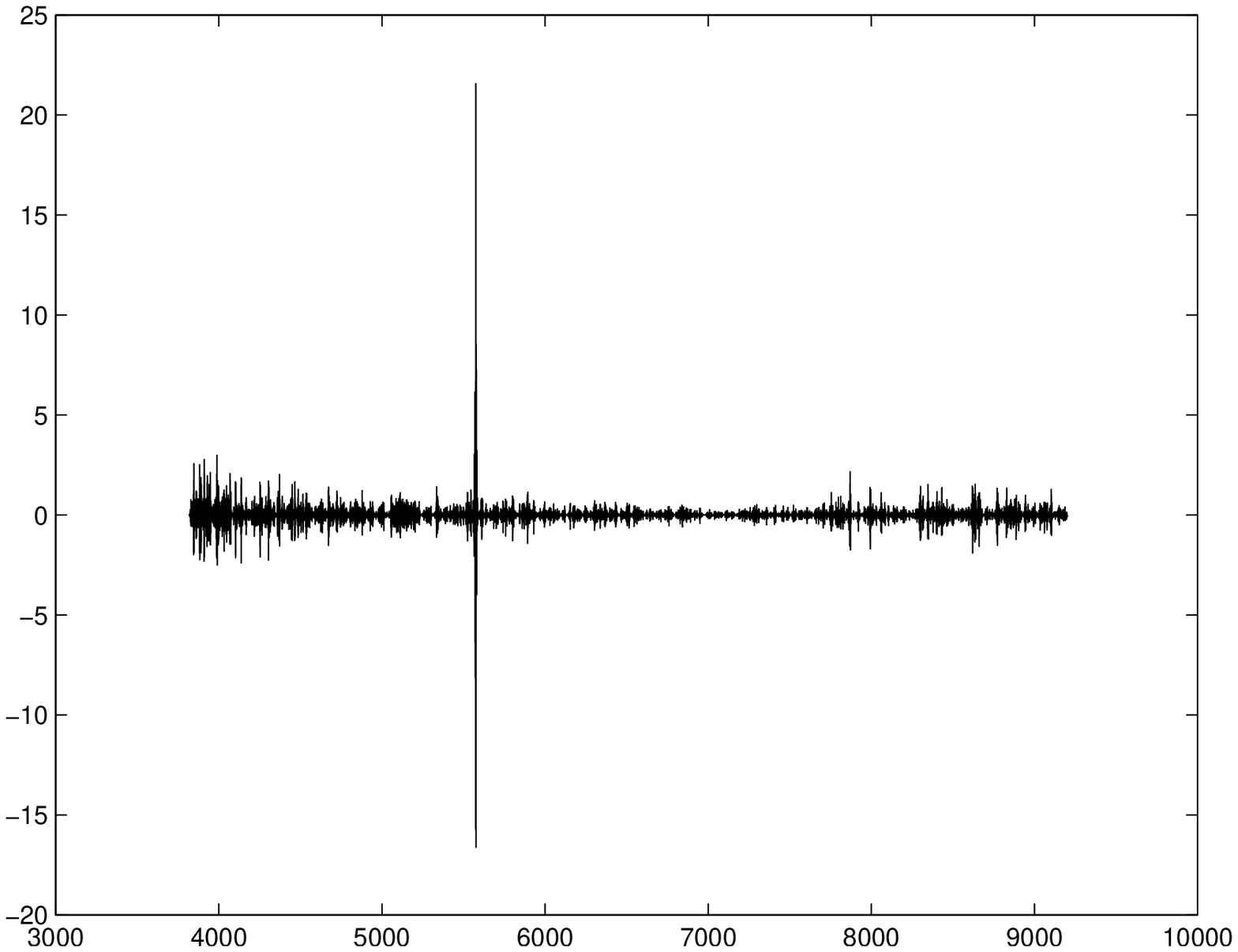}}
\hspace{-0.17in}
  \subfigure[WPD  $\bm{x}{[}2,3{]}$.]{
    \label{Fig:wp_dec:x_2_3} 
    \includegraphics[width =2.1in]{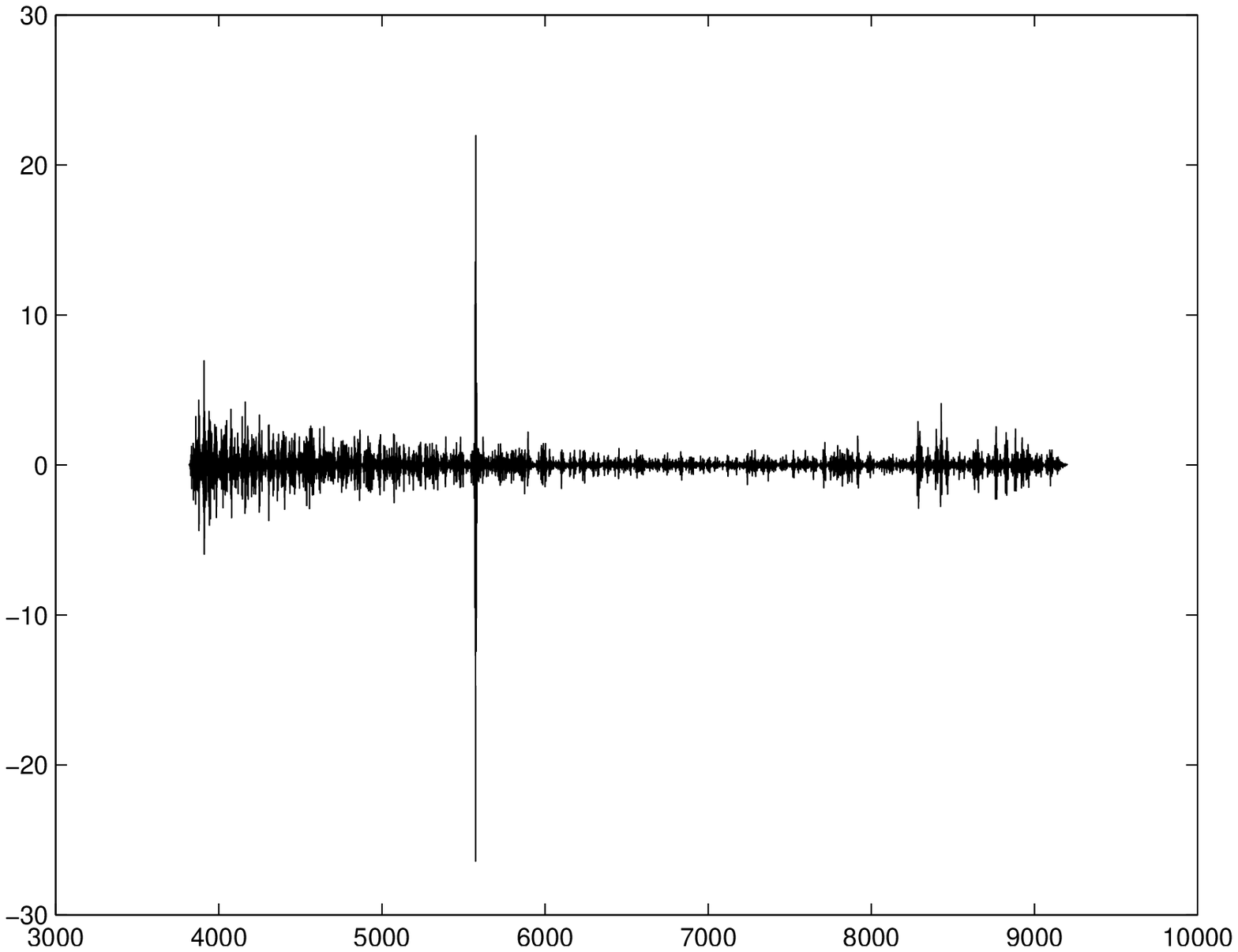}}
    \setlength{\abovecaptionskip}{-10pt}
  \caption{Spectrum and its WPD. This decomposition is computed based on wavelet basis rbio. WPD:Wavelet packet decomposition.}
  \label{Fig:wp_dec} 
\end{figure*}

Suppose that $j_0$ is an integer satisfying $0 \leq j_0 < 2^i$, $\bm{s}[i] = \{\bm{s}[i,j],j = 0, \cdots, 2^i -1 \}$, where $\bm{s}[i,j_0] = \bm{x}[i,j_0]$ and $\bm{s}[i,j] = \bm{0}$ \footnote{A zero vector sharing the same length with $\bm{x}[i,j]$.} if $j \neq j_0$ and $0 \leq j < 2^i$. Using WP reconstruction, we can map $\bm{s}$ to a vector $wprec(\bm{s}[i]) \in R^n$ to visualize the frequency sub-band $\bm{x}[i,j_0]$ in spectral space (Fig. \ref{Fig:wp_dec}). This visualizing technique is widely used in related research.

\subsubsection{Wavelength/Time-frequency Decomposition}
\begin{figure*}
\begin{center}
\includegraphics[ width =3in]{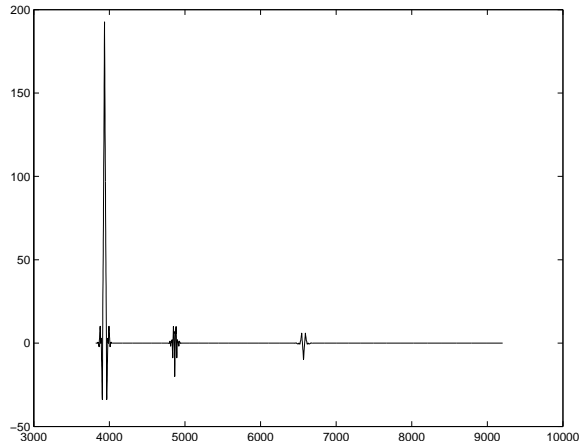}
\end{center}
\setlength{\abovecaptionskip}{1pt} \caption{Wavelength/time localization of wavelet packet decomposition. This is the visualization of three wavelet packet coefficients, $x_{5, 0}^9$, $x_{5, 1}^{78}$, $x_{5, 3}^{37}$ of the spectrum in Fig. \ref{Fig:wp_dec:x}. There are three areas with non-zero energy (non-zero areas). The three non-zero areas from left to right correspond to $x_{5, 0}^9$, $x_{5, 3}^{37}$, and $x_{5, 1}^{78}$, respectively.}
\label{Fig:T1_T14_T20}
\end{figure*}

A WPD coefficient can be visualized in the spectral space based on the method in Section \ref{Sec:Spectrum_Decomposition:Rec_Vis} (Fig. \ref{Fig:T1_T14_T20}). It can be shown that the energy of a WPD coefficient exists in a local and limited area (Fig. \ref{Fig:T1_T14_T20}), and a spectrum $\bm{x}$ can be reconstructed by the coefficients on a decomposition level $\bm{x}[i]$ (Section \ref{Sec:Spectrum_Decomposition:Rec_Vis} and \cite{Book:Mallat:2009}). Therefore, in addition to decomposing a signal-based frequency (Section \ref{Sec:Spectrum_Decomposition:wavelet_packet_transform}, Fig. \ref{Fig:wp_dec}), WP also implements wavelength decomposition. These characteristics are called wavelength/time--frequency localization or wavelength/time--frequency analysis \footnote{In the information processing community, a signal is  usually composed of some detected energy values on a series of time points, and thus, semantically, the above-mentioned characteristics of WPs are usually referred to as time--frequency localization or time--frequency analysis.}.

In this work, the wavelength position of a coefficient of WPD is represented by the center of the corresponding non-zero area in spectral space (Fig. \ref{Fig:T1_T14_T20}).

\begin{table*}\scriptsize
\setlength{\abovecaptionskip}{-100pt}
\setlength{\belowcaptionskip}{-100pt}
\centering
\caption{The Detected Typical Wavelength and Frequency for Estimating Atmospheric Parameters from Stellar Spectra }
\begin{tabular}{p{1cm}<{\centering} p{2.2cm}<{\centering} p{0.8cm}<{\centering}  p{1cm}<{\centering} p{2.2cm}<{\centering} p{0.8cm}<{\centering} p{1cm}<{\centering} p{2.2cm}<{\centering} p{0.8cm}<{\centering}}
  \hline \hline
\multicolumn{9}{c}{(a) The Detected Features for $T_\texttt{eff}$ Based on Basis Function rbio with the Optimal Decomposition Level 5 } \\
\hline
 Label    &TW $\lambda$(\AA)&IF&label    &TW $\lambda$(\AA)&IF&label    &TW $\lambda$(\AA)&IF \\
  T$_{1  }$&   [3825.6,3936.4,4050.4]   &   0   &T$_{2  }$&  [4118.1,4237.4,4360.1]   &   0   &T$_{3  }$&  [4633.4,4767.6,4905.7]   &   0 \\
  T$_{4  }$&   [4737.0,4874.2,5015.3]   &   0   &T$_{5  }$&  [4987.7,5132.2,5280.8]   &   0   &T$_{6  }$&  [5061.7,5208.3,5359.2]   &   0 \\
  T$_{7  }$&   [3818.6,3903.9,3991.2]   &   1   &T$_{8  }$&  [3998.5,4099.2,4202.4]   &   1   &T$_{9  }$&  [4241.3,4348.1,4457.6]   &   1 \\
  T$_{10 }$&   [4737.0,4856.2,4978.5]   &   1   &T$_{11 }$&  [4772.0,4892.2,5015.3]   &   1   &T$_{12 }$&  [5061.7,5189.2,5319.9]   &   1 \\
  T$_{13 }$&   [5099.2,5227.6,5359.2]   &   1   &T$_{14 }$&  [6407.7,6569.0,6734.4]   &   1   &T$_{15 }$&  [3839.7,3943.7,4050.4]   &   2 \\
  T$_{16 }$&   [5006.1,5141.6,5280.8]   &   2   &T$_{17 }$&  [5080.4,5218.0,5359.2]   &   2   &T$_{18 }$&  [5310.1,5453.8,5601.4]   &   2 \\
  T$_{19 }$&   [3818.6,3903.9,3991.2]   &   3   &T$_{20 }$&  [4754.4,4865.2,4978.5]   &   3   &T$_{21 }$&  [3818.6,3875.3,3932.8]   &   6 \\
  T$_{22 }$&   [3846.8,3947.3,4050.4]   &   6   &T$_{23 }$&  [3850.3,3934.6,4020.7]   &   15  &         &          \\
\hline
\multicolumn{9}{c}{(b) The Detected Features for log~$g$ Based on Basis Function coif with the Optimal Decomposition Level 6. } \\
\hline
  Label    &TW $\lambda$(\AA)&IF&label    &TW $\lambda$(\AA)&IF&label    &TW $\lambda$(\AA)&IF \\
  L$_{1  }$&   [3818.6,4264.8,4762.1]   &   0   &L$_{2  }$&  [3818.6,4360.1,4977.4]   &   0   &L$_{3  }$&  [3818.6,4392.4,5051.3]   &   0 \\
  L$_{4  }$&   [3818.6,4456.6,5202.4]   &   1   &L$_{5  }$&  [3894.9,4601.5,5437.5]   &   1   &L$_{6  }$&  [3952.8,4670.9,5518.2]   &   1 \\
  L$_{7  }$&   [4382.3,5178.5,6117.9]   &   1   &L$_{8  }$&  [5547.5,6553.9,7744.6]   &   1   &L$_{9  }$&  [5629.9,6652.7,7859.6]   &   1 \\
  L$_{10 }$&   [3818.6,3846.8,3874.4]   &   1   &L$_{11 }$&  [3818.6,4264.8,4762.1]   &   1   &L$_{12 }$&  [3952.8,4670.9,5518.2]   &   2 \\
  L$_{13 }$&   [4447.3,5255.3,6208.7]   &   2   &L$_{14 }$&  [4513.4,5333.3,6300.9]   &   2   &L$_{15 }$&  [5307.6,6271.9,7409.7]   &   2 \\
  L$_{16 }$&   [7449.0,8279.4,9202.4]   &   3   &L$_{17 }$&  [7559.6,8340.7,9202.4]   &   3   &L$_{18 }$&  [3818.6,4202.4,4623.8]   &   3 \\
  L$_{19 }$&   [4858.5,5741.2,6782.7]   &   3   &L$_{20 }$&  [7232.7,8158.3,9202.4]   &   3   &L$_{21 }$&  [7340.1,8218.6,9202.4]   &   4 \\
  L$_{22 }$&   [3818.6,3846.8,3874.4]   &   4   &L$_{23 }$&  [3818.6,4424.9,5126.3]   &   4   &L$_{24 }$&  [3894.9,4601.5,5437.5]   &   4 \\
  L$_{25 }$&   [7449.0,8279.4,9202.4]   &   5   &L$_{26 }$&  [7559.6,8340.7,9202.4]   &   5   &L$_{27 }$&  [3818.6,4424.9,5126.3]   &   5 \\
  L$_{28 }$&   [7126.9,8098.4,9202.4]   &   6   &L$_{29 }$&  [3818.6,4233.5,4692.5]   &   6   &L$_{30 }$&  [3818.6,4295.4,4832.8]   &   6 \\
  L$_{31 }$&   [7340.1,8218.6,9202.4]   &   6   &L$_{32 }$&  [3818.6,4171.6,4556.2]   &   7   &L$_{33 }$&  [3818.6,4392.4,5051.3]   &   8 \\
  L$_{34 }$&   [4255.0,5028.1,5940.2]   &   8   &L$_{35 }$&  [5003.8,5911.5,6985.5]   &   8   &L$_{36 }$&  [6919.9,7979.9,9202.4]   &   8 \\
  L$_{37 }$&   [7126.9,8098.4,9202.4]   &   8   &L$_{38 }$&  [7340.1,8218.6,9202.4]   &   9   &L$_{39 }$&  [3818.6,4392.4,5051.3]   &   9 \\
  L$_{40 }$&   [4071.0,4810.6,5683.3]   &   9   &L$_{41 }$&  [4513.4,5333.3,6300.9]   &   9   &L$_{42 }$&  [3818.6,4140.0,4489.5]   &   9 \\
  L$_{43 }$&   [6241.7,7375.6,8713.6]   &   10  &L$_{44 }$&  [3818.6,4295.4,4832.8]   &   11  &L$_{45 }$&  [4131.4,4882.0,5767.7]   &   11\\
  L$_{46 }$&   [4255.0,5028.1,5940.2]   &   11  &L$_{47 }$&  [7340.1,8218.6,9202.4]   &   12  &L$_{48 }$&  [4192.8,4954.5,5853.3]   &   12\\
  L$_{49 }$&   [4382.3,5178.5,6117.9]   &   12  &L$_{50 }$&  [7340.1,8218.6,9202.4]   &   12  &L$_{51 }$&  [7449.0,8279.4,9202.4]   &   16\\
  L$_{52 }$&   [3818.6,4328.1,4904.6]   &   19  &L$_{53 }$&  [3818.6,4490.6,5279.6]   &   20  &L$_{54 }$&  [4192.8,4954.5,5853.3]   &   20\\
  L$_{55 }$&   [7022.6,8039.0,9202.4]   &   20  &L$_{56 }$&  [4513.4,5333.3,6300.9]   &   21  &L$_{57 }$&  [3818.6,4050.4,4295.4]   &   21\\
  L$_{58 }$&   [4580.4,5412.5,6394.4]   &   21  &L$_{59 }$&  [3818.6,4424.9,5126.3]   &   21  &L$_{60 }$&  [4255.0,5028.1,5940.2]   &   24\\
  L$_{61 }$&   [3818.6,4456.6,5202.4]   &   25  &L$_{62 }$&  [4318.2,5101.5,6028.4]   &   25  &         &          \\
\hline
\multicolumn{9}{c}{(c) The Detected Features for [Fe/H] Based on Basis Function rbio with the Optimal Decomposition Level 4. } \\
\hline
 Label    &TW $\lambda$(\AA)&IF&label    &TW $\lambda$(\AA)&IF&label    &TW $\lambda$(\AA)&IF \\
 F$_{1  }$&   [3882.4,3936.4,3991.2]   &   0   &F$_{2  }$&  [4449.4,4511.3,4574.0]   &   0   &F$_{3  }$&  [4565.6,4629.1,4693.5]   &   0 \\
  F$_{4  }$&   [4684.9,4750.1,4816.1]   &   0   &F$_{5  }$&  [4789.6,4856.2,4923.8]   &   0   &F$_{6  }$&  [4807.3,4874.2,4942.0]   &   0 \\
  F$_{7  }$&   [3896.7,3943.7,3991.2]   &   1   &F$_{8  }$&  [4043.0,4091.7,4141.0]   &   1   &F$_{9  }$&  [4118.1,4167.7,4217.9]   &   1 \\
  F$_{10 }$&   [4498.8,4553.0,4607.9]   &   1   &F$_{11 }$&  [4532.1,4586.7,4641.9]   &   1   &F$_{12 }$&  [4616.4,4672.0,4728.2]   &   1 \\
  F$_{13 }$&   [4650.5,4706.5,4763.2]   &   1   &F$_{14 }$&  [4667.7,4723.9,4780.8]   &   1   &F$_{15 }$&  [4719.5,4776.4,4833.9]   &   1 \\
  F$_{16 }$&   [4789.6,4847.3,4905.7]   &   1   &F$_{17 }$&  [4825.0,4883.1,4942.0]   &   1   &F$_{18 }$&  [4842.8,4901.2,4960.2]   &   1 \\
  F$_{19 }$&   [4932.9,4992.3,5052.4]   &   1   &F$_{20 }$&  [4969.4,5029.2,5089.8]   &   1   &F$_{21 }$&  [5024.6,5085.1,5146.4]   &   1 \\
  F$_{22 }$&   [5043.1,5103.9,5165.4]   &   1   &F$_{23 }$&  [5061.7,5122.7,5184.4]   &   1   &F$_{24 }$&  [5080.4,5141.6,5203.6]   &   1 \\
  F$_{25 }$&   [5099.2,5160.6,5222.8]   &   1   &F$_{26 }$&  [5118.0,5179.6,5242.0]   &   1   &F$_{27 }$&  [5174.9,5237.2,5300.3]   &   1 \\
  F$_{28 }$&   [5194.0,5256.5,5319.9]   &   1   &F$_{29 }$&  [5290.5,5354.3,5418.8]   &   1   &F$_{30 }$&  [5591.1,5658.5,5726.6]   &   1 \\
  F$_{31 }$&   [5632.5,5700.3,5769.0]   &   1   &F$_{32 }$&  [8415.9,8517.3,8619.9]   &   1   &F$_{33 }$&  [8447.0,8548.7,8651.7]   &   1 \\
  F$_{34 }$&   [4424.9,4482.3,4540.5]   &   2   &F$_{35 }$&  [4507.1,4565.6,4624.9]   &   2   &F$_{36 }$&  [4960.2,5024.6,5089.8]   &   2 \\
  F$_{37 }$&   [4978.5,5043.1,5108.6]   &   2   &F$_{38 }$&  [5033.8,5099.2,5165.4]   &   2   &F$_{39 }$&  [5146.4,5213.1,5280.8]   &   2 \\
  F$_{40 }$&   [5203.6,5271.1,5339.5]   &   2   &F$_{41 }$&  [5319.9,5388.9,5458.8]   &   2   &F$_{42 }$&  [3889.6,3932.8,3976.5]   &   3 \\
  F$_{43 }$&   [4005.9,4050.4,4095.4]   &   3   &F$_{44 }$&  [4202.4,4249.1,4296.4]   &   3   &F$_{45 }$&  [4249.1,4296.4,4344.1]   &   3 \\
  F$_{46 }$&   [4344.1,4392.4,4441.2]   &   3   &F$_{47 }$&  [4408.6,4457.6,4507.1]   &   3   &F$_{48 }$&  [4424.9,4474.0,4523.8]   &   3 \\
  F$_{49 }$&   [4490.6,4540.5,4590.9]   &   3   &F$_{50 }$&  [4798.4,4851.8,4905.7]   &   3   &F$_{51 }$&  [4869.7,4923.8,4978.5]   &   3 \\
  F$_{52 }$&   [4923.8,4978.5,5033.8]   &   3   &F$_{53 }$&  [4942.0,4996.9,5052.4]   &   3   &F$_{54 }$&  [5033.8,5089.8,5146.4]   &   3 \\
  F$_{55 }$&   [5300.3,5359.2,5418.8]   &   3   &F$_{56 }$&  [5418.8,5479.0,5539.9]   &   3   &F$_{57 }$&  [5580.8,5642.9,5705.6]   &   3 \\
  F$_{58 }$&   [7689.5,7775.0,7861.4]   &   3   &F$_{59 }$&  [5103.9,5172.5,5242.0]   &   4   &F$_{60 }$&  [4175.4,4227.7,4280.6]   &   6 \\
  F$_{61 }$&   [4364.2,4418.8,4474.0]   &   6   &F$_{62 }$&  [4494.7,4550.9,4607.9]   &   6   &F$_{63 }$&  [4856.2,4917.0,4978.5]   &   6 \\
  F$_{64 }$&   [5246.9,5312.5,5379.0]   &   6   &F$_{65 }$&  [4129.5,4173.5,4217.9]   &   7   &F$_{66 }$&  [4284.5,4330.1,4376.2]   &   7 \\
  F$_{67 }$&   [5132.2,5186.8,5242.0]   &   7   &F$_{68 }$&  [5305.2,5361.7,5418.8]   &   7   &         &          \\
\hline
\\
\multicolumn{9}{p{15cm}}{Note. TW: typical wavelength position represented by a three-dimensional vector [a, b, c], where a, b, c are respectively the starting wavelength, central wavelength, ending wavelength, and log$_{10}$b = (log$_{10}$a + log$_{10}$c)/2.
IF: Index of sub-bands with different frequencies. In (a), 0.56\% of the 4096 wavelet packet components/coefficients are extracted to estimate $T_\texttt{eff}$; (b) 1.18\% of the 5248 wavelet packet components/coefficients are extracted to estimate log~$g$; (c) 1.72\% of the 3952 wavelet packet components/coefficients are extracted to estimate [Fe/H]. Selection of basis function and the decomposition level are discussed in Section \ref{Sec:More:Technical:Config_wavelet_Decom}.}

\end{tabular}\label{Tab:Detected_features}
\end{table*}

\subsection{Feature Selection}\label{Sec:Spectrum_Decomposition:feature_selection}
This subsection focuses on selecting a linearly supporting subset of WP components/coefficients to estimate the atmospheric parameters $T_\texttt{eff}$, log$~g$, and [Fe/H]. This process is referred to as a feature selection problem in machine learning.

The high-frequency WP-decomposed components usually have a larger probability of being affected by noise than the low-frequency components. In the literature, therefore, features are usually selected by throwing away as noise those components with frequencies larger than the preset threshold \citep{Journal:Lu:2013}. Assessing this threshold is subjective. Furthermore, apart from noise, there exists a high level of redundancy in a stellar spectrum when estimating atmospheric parameters \citep{Journal:Li:2014}.

Therefore, we analyze the correlation between WP components and the atmospheric parameters to be estimated and detect representative spectral features using LASSO(LARS)$_{\texttt{bs}}$ (Section \ref{Sec:sub:Model_Further}). The WP components selected as useful features are presented in Table \ref{Tab:Detected_features} and Fig. \ref{Fig:feature:visualization}, and more technical details are discussed in Section \ref{Sec:More:Technical}.

\emph{Visualization of the features.}~~~~Based on the results in Table \ref{Tab:Detected_features}, a spectrum $\bm{x}$ should be decomposed to the fifth-level $\bm{x}[5]$ to estimate $T_{\texttt{eff}}$. A vector $\bm{s}[5]$ can be constructed from $\bm{x}[5]$ by keeping the elements of $\bm{x}[5]$ corresponding to the features in Table \ref{Tab:Detected_features} (a) but reset the other elements of $\bm{x}[5]$ to zero. Thus, we can visualize the features of $\bm{x}$ in a spectral space through WP reconstruction $wprec(\bm{s}[5])$ (Section \ref{Sec:Spectrum_Decomposition:Rec_Vis}, Fig. \ref{Fig:feature:visualization:Teff2}). Similarly, the features in Tables \ref{Tab:Detected_features} (b) and (c) can also be visualized in the spectral space (Fig. \ref{Fig:feature:visualization:logg2}, Fig. \ref{Fig:feature:visualization:feh2}) through WP reconstruction.

We find that the extracted features are a subset of WP components/coefficients in some lower sub-bands (Table \ref{Tab:Detected_features}, Fig. \ref{Fig:feature:distribution}). In other words, not only are some of the sub-bands with higher frequency are ineffective but also many components in the sub-bands with lower frequency appear redundant. Further discussion and the corresponding results are presented in Section \ref{Sec:More:Technical}.

To estimate $T_\texttt{eff}$, a spectrum is decomposed into $2^5 = 32$ sub-bands with a frequency index from 0 to 31; there exist 120 components in each sub-band, and the detected features come only from sub-bands 0, 1, 2, 3, 6, and 16: this means that 99.33\% of the components/coefficients are redundancy or noise in each sub-band (Fig. \ref{Fig:feature:distribution:Teff}). Similarly, to estimate log~$g$, a spectrum is decomposed into 64 sub-bands, more than 85\% of the WP components are redundancy and noise in each sub-band (Fig. \ref{Fig:feature:distribution:logg}). To estimate [Fe/H], a spectrum is decomposed into 16 sub-bands, and more than 88.7\% of the WP components are redundancy and noise in each sub-band (Fig. \ref{Fig:feature:distribution:FeH}). An interesting phenomenon is that all components in the lowest frequency sub-band are redundancy or noise when estimating log~$g$. Therefore, the effectiveness of a WP component depends both on its frequency and on its wavelength (Table \ref{Tab:Detected_features}).

Based on the detected features in Table \ref{Tab:Detected_features}, three atmospheric parameter estimate models can be established using OLS (Equations (\ref{Equ:MSE}) and (\ref{Equ:linear:model})). The coefficients of these models are given in Table \ref{Tab:coefficients}. They quantify the association between the detected spectral features and the atmospheric parameter to be estimated. As already mentioned in Section \ref{Sec:Estimation_LASSO:Model}, these coefficients can be interpreted as the average effect of a one-unit increase in a spectral feature \citep{Book:James:2013}. For example, the coefficient $w_1$ is 0.2379 in the $T_\texttt{eff}$ estimate model (Table \ref{Tab:coefficients}); therefore, if the spectral feature T$_1$ increases one unit with all other features $\{ T_j, j= 2,\cdots, 23\}$ remaining fixed, then the effective temperature log$~T_\texttt{eff}$ will increase 0.2379.

\begin{table*}\scriptsize
\setlength{\abovecaptionskip}{-100pt}
\setlength{\belowcaptionskip}{-100pt}
\centering
\caption{Coefficients of the Atmospheric Parameter Estimation Model learned by OLS from SDSS spectra (More Details of the Experiment are Presented in Section \ref{Sec:estimation:SDSS})}
\begin{tabular}{cccccccccccccc}
  \hline \hline
\multicolumn{14}{c}{(a) Detected Features for $T_\texttt{eff}$ Based on Wavelet Basis Function rbio with the Optimal Decomposition Level 5. } \\
\hline
  Label   &    $w_i$       &label    &    $w_i$       & label  &    $w_i$     &  label  &    $w_i$        &label   &    $w_i$       & label  &    $w_i$      & label  &    $w_i$       \\
  T$_{1  }$&    0.2379  &T$_{2  }$&    0.4505  &T$_{3  }$&    0.4819  &T$_{4  }$&   -1.8345  &T$_{5  }$&    1.0740  &T$_{6  }$&    0.7236  &T$_{7  }$&    0.3264\\
  T$_{8  }$&    0.4558  &T$_{9  }$&    0.7899  &T$_{10 }$&    1.0640  &T$_{11 }$&   -0.9495  &T$_{12 }$&    0.8358  &T$_{13 }$&    0.8650  &T$_{14 }$&    1.5004\\
  T$_{15 }$&    0.2825  &T$_{16 }$&    0.5646  &T$_{17 }$&    0.5911  &T$_{18 }$&   -0.4539  &T$_{19 }$&    0.3262  &T$_{20 }$&   -0.9225  &T$_{21 }$&   -0.2100\\
  T$_{22 }$&    0.2366  &T$_{23 }$&    0.4507  &         &            &         &            &         &            &         &            &         &          \\

\hline
\multicolumn{14}{c}{(b) Detected Features for log~$g$ Based on Wavelet Basis Function coif with the Optimal Decomposition Level 6. } \\
\hline
  Label   &    $w_i$       &label    &    $w_i$       & label  &    $w_i$     &  label  &    $w_i$        &label   &    $w_i$       & label  &    $w_i$      & label  &    $w_i$       \\
  L$_{1  }$&   -14.9500  &L$_{2  }$&    11.9097  &L$_{3  }$&   -15.9727  &L$_{4  }$&   -29.8595  &L$_{5  }$&    20.9904  &L$_{6  }$&   -12.1664  &L$_{7  }$&    48.9051\\
  L$_{8  }$&    124.5594  &L$_{9  }$&   -23.4959  &L$_{10 }$&    4.2752  &L$_{11 }$&    9.3749  &L$_{12 }$&   -11.0132  &L$_{13 }$&   -26.4751  &L$_{14 }$&   -35.4924\\
  L$_{15 }$&    20.5712  &L$_{16 }$&    26.5010  &L$_{17 }$&    32.2221  &L$_{18 }$&    11.0103  &L$_{19 }$&   -22.3564  &L$_{20 }$&    11.6441  &L$_{21 }$&   -14.6373\\
  L$_{22 }$&   -6.9629  &L$_{23 }$&   -10.0933  &L$_{24 }$&   -11.4390  &L$_{25 }$&   -26.8052  &L$_{26 }$&   -25.4320  &L$_{27 }$&    9.7027  &L$_{28 }$&   -21.6698\\
  L$_{29 }$&    16.4676  &L$_{30 }$&   -9.1541  &L$_{31 }$&   -47.8560  &L$_{32 }$&   -7.0705  &L$_{33 }$&   -18.4181  &L$_{34 }$&   -19.0691  &L$_{35 }$&   -19.7537\\
  L$_{36 }$&   -43.4952  &L$_{37 }$&   -11.2162  &L$_{38 }$&   -12.9840  &L$_{39 }$&    8.6831  &L$_{40 }$&    13.4435  &L$_{41 }$&   -13.0983  &L$_{42 }$&   -5.3274\\
  L$_{43 }$&   -19.2899  &L$_{44 }$&   -9.5705  &L$_{45 }$&    7.2348  &L$_{46 }$&   -12.2654  &L$_{47 }$&    16.4605  &L$_{48 }$&   -14.0761  &L$_{49 }$&    34.1498\\
  L$_{50 }$&   -16.5275  &L$_{51 }$&   -8.4650  &L$_{52 }$&   -14.5751  &L$_{53 }$&   -19.9664  &L$_{54 }$&   -28.4143  &L$_{55 }$&   -17.2212  &L$_{56 }$&    18.7294\\
  L$_{57 }$&   -3.6372  &L$_{58 }$&   -28.5637  &L$_{59 }$&   -8.3460  &L$_{60 }$&    38.7862  &L$_{61 }$&   -15.1417  &L$_{62 }$&   -23.3872  &         &          \\

\hline
\multicolumn{14}{c}{(c) Detected Features for [Fe/H] Based on Wavelet Basis Function rbio with the Optimal Decomposition Level 4. } \\
\hline
  Label   &    $w_i$       &label    &    $w_i$       & label  &    $w_i$     &  label  &    $w_i$        &label   &    $w_i$       & label  &    $w_i$      & label  &    $w_i$       \\
  F$_{1  }$&   -15.9393  &F$_{2  }$&    14.4512  &F$_{3  }$&    10.0133  &F$_{4  }$&    14.5021  &F$_{5  }$&   -36.4275  &F$_{6  }$&   -22.7435  &F$_{7  }$&   -20.0956\\
  F$_{8  }$&   -3.5740  &F$_{9  }$&   -4.2593  &F$_{10 }$&    15.7326  &F$_{11 }$&    15.0760  &F$_{12 }$&    9.5922  &F$_{13 }$&    15.1001  &F$_{14 }$&   -10.5834\\
  F$_{15 }$&   -16.4978  &F$_{16 }$&   -19.8914  &F$_{17 }$&   -6.1936  &F$_{18 }$&   -13.9105  &F$_{19 }$&   -9.9439  &F$_{20 }$&   -16.1353  &F$_{21 }$&    10.6011\\
  F$_{22 }$&    14.8693  &F$_{23 }$&   -6.8806  &F$_{24 }$&    14.3556  &F$_{25 }$&   -19.2167  &F$_{26 }$&   -43.5510  &F$_{27 }$&    3.8668  &F$_{28 }$&   -19.1643\\
  F$_{29 }$&   -7.0540  &F$_{30 }$&    13.4940  &F$_{31 }$&   -13.5002  &F$_{32 }$&   -8.6885  &F$_{33 }$&    9.5827  &F$_{34 }$&    17.6601  &F$_{35 }$&    5.7138\\
  F$_{36 }$&   -12.3663  &F$_{37 }$&    5.2848  &F$_{38 }$&    6.5740  &F$_{39 }$&   -21.8866  &F$_{40 }$&   -23.7674  &F$_{41 }$&   -16.2051  &F$_{42 }$&   -10.6150\\
  F$_{43 }$&    4.5351  &F$_{44 }$&   -5.7992  &F$_{45 }$&    7.4311  &F$_{46 }$&    1.2759  &F$_{47 }$&   -8.2072  &F$_{48 }$&    5.1707  &F$_{49 }$&    1.6828\\
  F$_{50 }$&    13.3047  &F$_{51 }$&   -11.3030  &F$_{52 }$&    8.5673  &F$_{53 }$&    7.0954  &F$_{54 }$&    6.1268  &F$_{55 }$&    15.5904  &F$_{56 }$&   -5.3280\\
  F$_{57 }$&   -16.7121  &F$_{58 }$&   -10.0843  &F$_{59 }$&   -17.2424  &F$_{60 }$&   -2.3355  &F$_{61 }$&   -4.8779  &F$_{62 }$&   -8.4915  &F$_{63 }$&    10.0789\\
  F$_{64 }$&    15.6214  &F$_{65 }$&    4.5253  &F$_{66 }$&   -6.8905  &F$_{67 }$&    17.1603  &F$_{68 }$&   -8.5783  &         &            &         &          \\
\hline
\\
\multicolumn{14}{p{16cm}}{
Note. More details of the experiment are presented in Section \ref{Sec:estimation:SDSS}. The coefficients predict the average effect of the corresponding spectral feature on the atmospheric parameter to be estimated. The labels of spectral features are defined in Table \ref{Tab:Detected_features}.}
\end{tabular}\label{Tab:coefficients}
\end{table*}

\subsection{Characteristic - Good Interpretability}\label{Sec:Spectrum_Decomposition:characteristic}
Due to the characteristic of the time--frequency localization of a wavelet basis function, every detected feature has a specific wavelength position (Table \ref{Tab:Detected_features}, Fig.~\ref{Fig:feature:distribution:Teff_2Dimage}, Fig.~\ref{Fig:feature:distribution:logg_2Dimage} and Fig.~\ref{Fig:feature:distribution:FeH_2Dimage}), which helps to trace back the physical effective factors and evaluate their contributions to the atmospheric parameter estimate from stellar spectra (Table \ref{Tab:coefficients}). For example,
 H$_\gamma$ is a sensitive line to surface temperature (T$_9$ in Tables \ref{Tab:Detected_features} and \ref{Tab:coefficients}),
 H$_\alpha$ is sensitive to both surface temperature and gravity (T$_{14}$ and L$_8$ in Tables \ref{Tab:Detected_features} and \ref{Tab:coefficients}),
 CaII$~K$ is sensitive to both surface temperature and metallicity (T$_1$, T$_{23}$, F$_{42}$ in Tables \ref{Tab:Detected_features} and \ref{Tab:coefficients}),
 and H$_\delta$ is sensitive to both surface temperature and metallicity (T$_8$ and F$_8$ in Tables \ref{Tab:Detected_features} and \ref{Tab:coefficients}).

Note, however, that the selected features in Table \ref{Tab:Detected_features} may span a somewhat larger wavelength width than traditionally used for stellar absorption lines. Thus, it could be asked whether some selected features may physically correspond to spectral blends rather than to single lines, which would explain why some wavelength-identified features unexpectedly appear sensitive to an atmospheric parameter: this is the case in H$_\delta$ for example, which should not, by itself and considered alone, be sensitive to metallicity. We also underline that the present study does not take into account the effects of spectral resolution on the effectiveness of the feature selection.

\subsection{Physical Dependence of the Detected Features and Their Contributions}\label{Sec:Spectrum_Decomposition:dependence}
The detected features and their contributions depend on the range of atmospheric parameters to be investigated. The following examples pertain to the effective temperature determination. Let us split the SDSS training spectra set (10,000 spectra) into the 4 following subsets based on the effective temperature derived by the SDSS SSPP:\\
\indent $S_1$: the spectra with $T_\texttt{eff} < 5200~K$,\\
\indent $S_2$: the spectra with $5200~K \leq T_\texttt{eff} < 6000~K$,\\
\indent $S_3$: the spectra with $6000~K \leq T_\texttt{eff} < 7500~K$,\\
\indent $S_4$: the spectra with $T_\texttt{eff} \geq 7500~K$.

Based on the features selected by LASSO(LARS)$_{\texttt{bs}}$ in Table \ref{Tab:Detected_features} (a), we are led to four models $M_1$, $M_2$, $M_3$, and $M_4$ corresponding to the four training subsets $S_1$, $S_2$, $S_3$, and $S_4$. The coefficients of these four models are presented in Table \ref{Tab:coefficients:dependences}. It is obvious that the contribution/coefficient of each detected feature T$_i$ depends on the range of effective temperatures: for example, the feature associated with the CaII K line (T$_1$, T$_{23}$) is only weakly pertinent for stars hotter than 6000 K.

 \begin{table*}\scriptsize
\setlength{\abovecaptionskip}{-100pt}
\setlength{\belowcaptionskip}{-100pt}
\centering
\caption{Dependences of the Contribution of the Detected Features on the Range of Effective Temperature}
\begin{tabular}{cccccccccc}
  \hline \hline

  Label   &    $w_i(M_1)$       &$w_i(M_2)$    &    $w_i(M_3)$       & $w_i(M_4)$  &    label   &    $w_i(M_1)$       &$w_i(M_2)$    &    $w_i(M_3)$       & $w_i(M_4)$  \\
  \hline
  T$_{1  }$&   0.2899  &  0.0951  &  0.0005   &   0.0672  &T$_{2  }$&   0.1103  &  0.3226  &  0.3159   &   0.0838\\
  T$_{3  }$&   0.9069  &  0.8461  &  0.6738   &   0.6006  &T$_{4  }$&   -0.7189  &  -2.6379  &  -2.1906   &   -1.1374\\
  T$_{5  }$&   -0.5341  &  0.8141  &  0.8674   &   1.3047  &T$_{6  }$&   -0.4443  &  1.0845  &  0.6762   &   0.1175\\
  T$_{7  }$&   -0.2309  &  0.2490  &  0.4674   &   -0.1354  &T$_{8  }$&   0.3543  &  0.4148  &  0.7816   &   0.5406\\
  T$_{9  }$&   0.6276  &  0.7579  &  0.9733   &   0.1057  &T$_{10 }$&   0.9216  &  1.1588  &  1.3794   &   -0.0292\\
  T$_{11 }$&   -3.0122  &  -0.5011  &  -0.5016   &   -0.4103  &T$_{12 }$&   0.5619  &  -0.0543  &  0.6089   &   2.1177\\
  T$_{13 }$&   -0.0779  &  0.1908  &  0.7772   &   1.3919  &T$_{14 }$&   -1.3688  &  0.7778  &  2.3492   &   1.2935\\
  T$_{15 }$&   0.8306  &  0.1825  &  0.2503   &   0.1015  &T$_{16 }$&   -0.5552  &  0.0912  &  0.7270   &   1.2507\\
  T$_{17 }$&   -0.9458  &  0.4172  &  0.1666   &   0.4198  &T$_{18 }$&   0.6194  &  -0.1032  &  -0.8771   &   -0.0588\\
  T$_{19 }$&   0.0948  &  0.1530  &  0.2582   &   0.0372  &T$_{20 }$&   1.2295  &  -0.3203  &  -2.1724   &   -0.7095\\
  T$_{21 }$&   -0.4782  &  -0.0629  &  -0.0731   &   -0.0845  &T$_{22 }$&   0.7560  &  -0.0193  &  0.0556   &   -0.0828\\
  T$_{23 }$&   0.8643  &  -0.1538  &  0.0331   &   0.0215  &         &          \\
 \hline
 \\
 \multicolumn{10}{p{13cm}}{Note. $M_1$, $M_2$, $M_3$, and $M_4$ are defined in Section \ref{Sec:Spectrum_Decomposition:dependence}. The labels of the spectral features are defined in Table \ref{Tab:Detected_features}. $w_i(M_j)$ represents the coefficient of model $M_j$.}
\end{tabular}\label{Tab:coefficients:dependences}
\end{table*}

\begin{figure*}
  \centering
\subfigure[Five spectra with different $T_{\texttt{eff}}$.]{
    \label{Fig:feature:visualization:Teff1} 
    \includegraphics[width =3.0in]{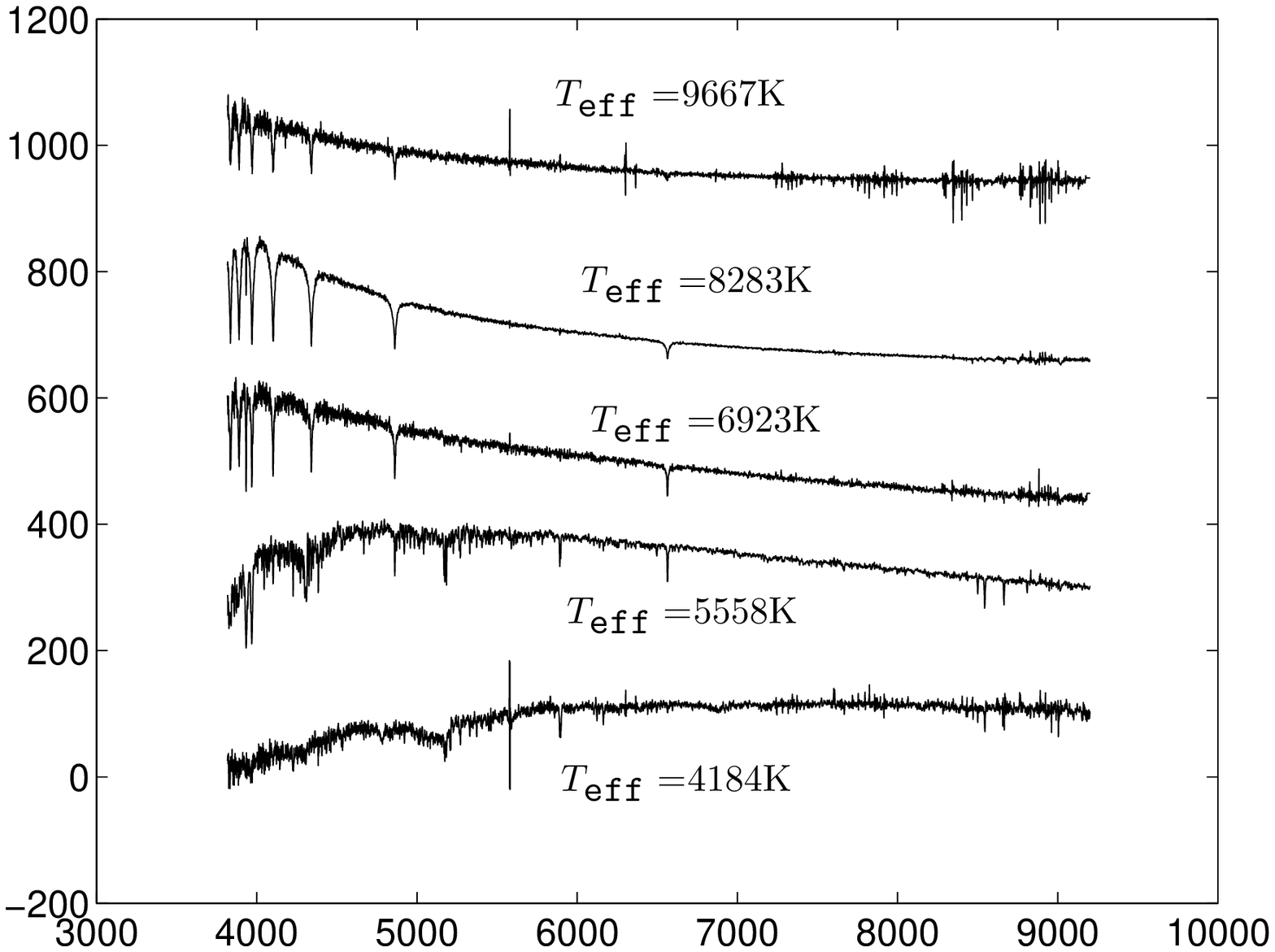}}
\hspace{-0.17in}
  \subfigure[Features of the spectra in Fig. \ref{Fig:feature:visualization:Teff1}.]{
    \label{Fig:feature:visualization:Teff2} 
    \includegraphics[width =3.0in]{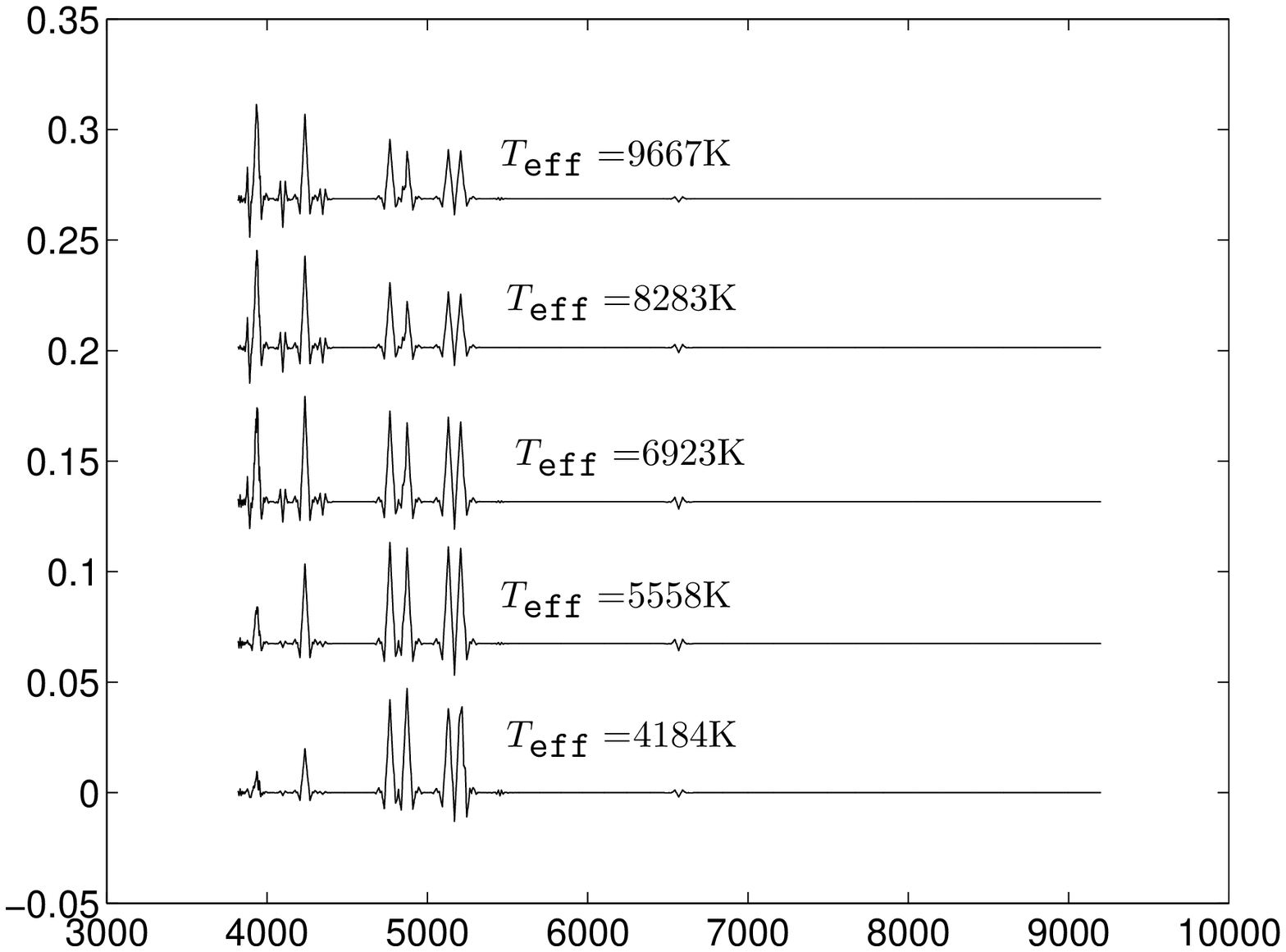}}
\hspace{-0.17in}
  \subfigure[Five spectra with different log$~g$.]{
    \label{Fig:feature:visualization:logg1} 
    \includegraphics[width =3.0in]{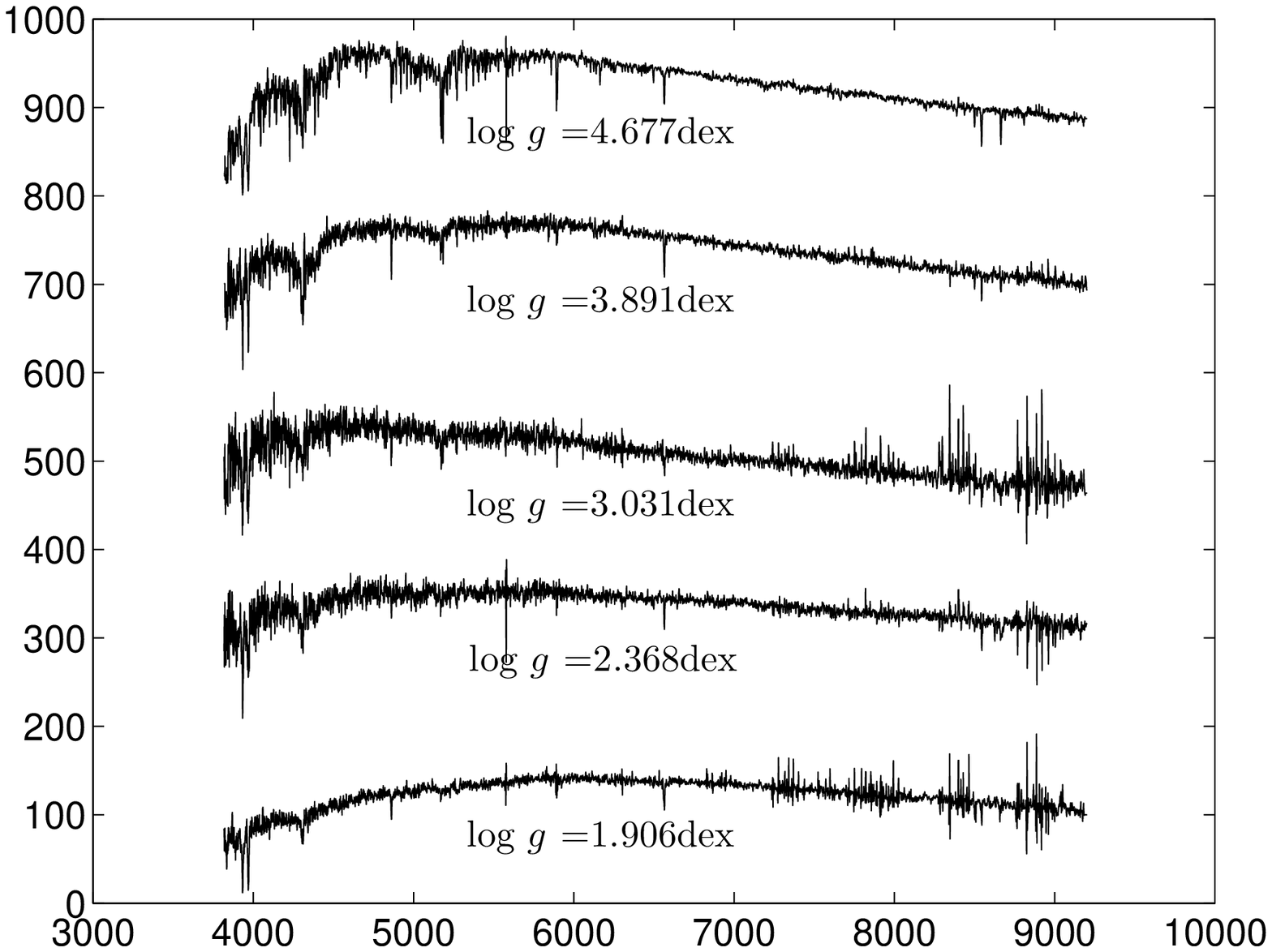}}
\hspace{-0.17in}
  \subfigure[Features of the spectra in Fig. \ref{Fig:feature:visualization:logg1}.]{
    \label{Fig:feature:visualization:logg2} 
    \includegraphics[width =3.0in]{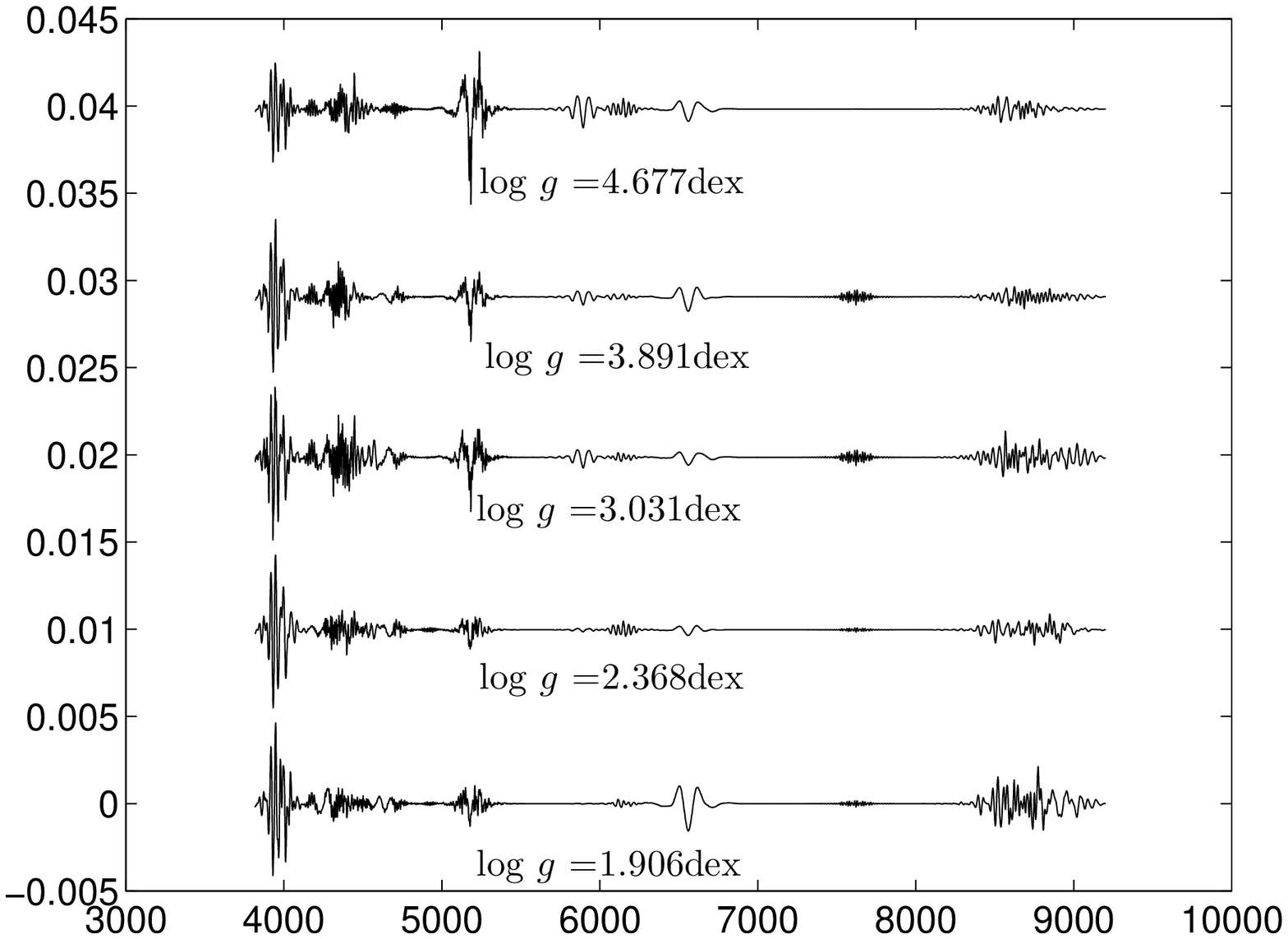}}
\hspace{-0.17in}
  \subfigure[Five spectra with different $\texttt{[}$Fe/H$\texttt{]}$.]{
    \label{Fig:feature:visualization:feh1} 
    \includegraphics[width =3.0in]{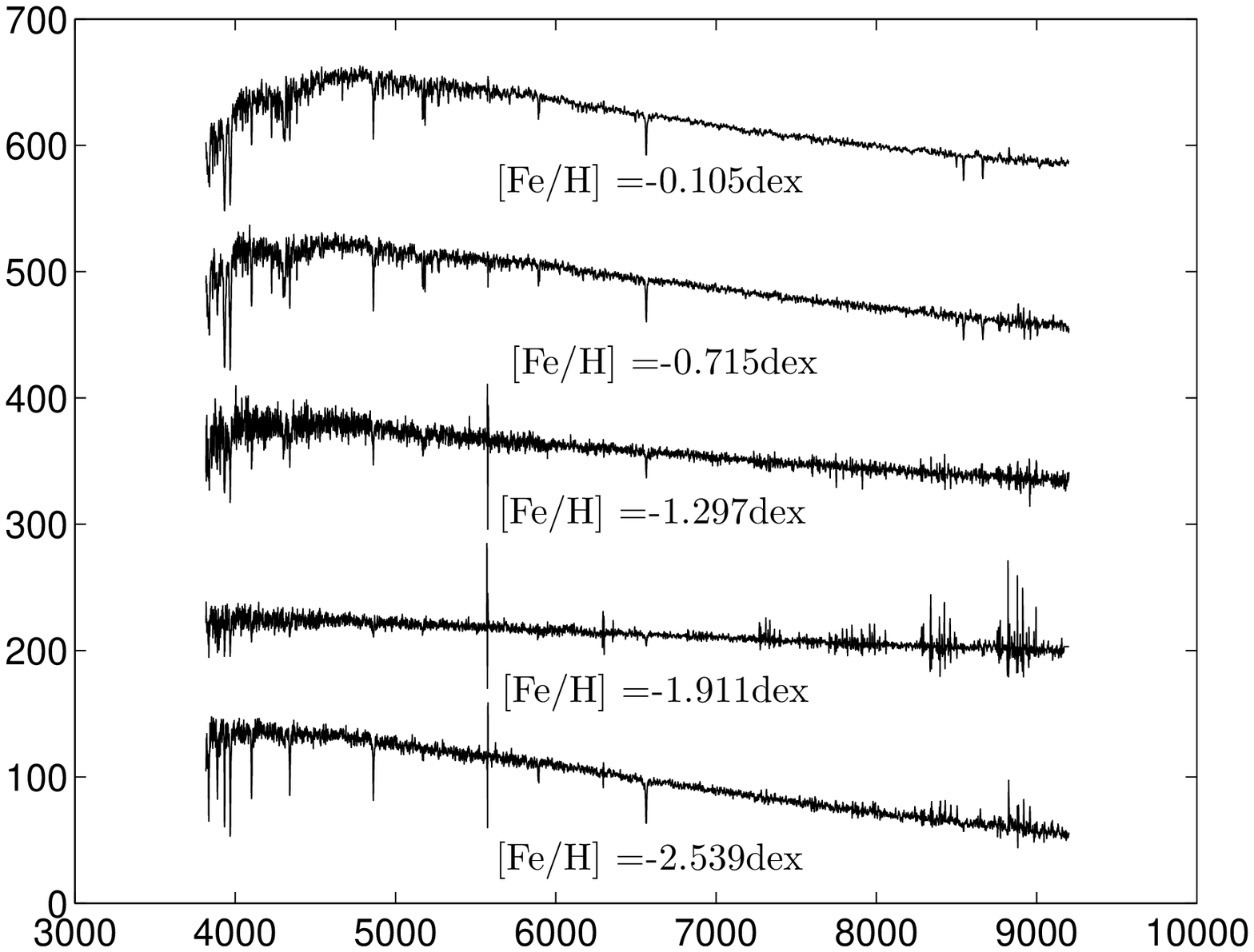}}
\hspace{-0.17in}
    \subfigure[Features of the spectra in Fig. \ref{Fig:feature:visualization:feh1}.]{
    \label{Fig:feature:visualization:feh2} 
    \includegraphics[width =3.0in]{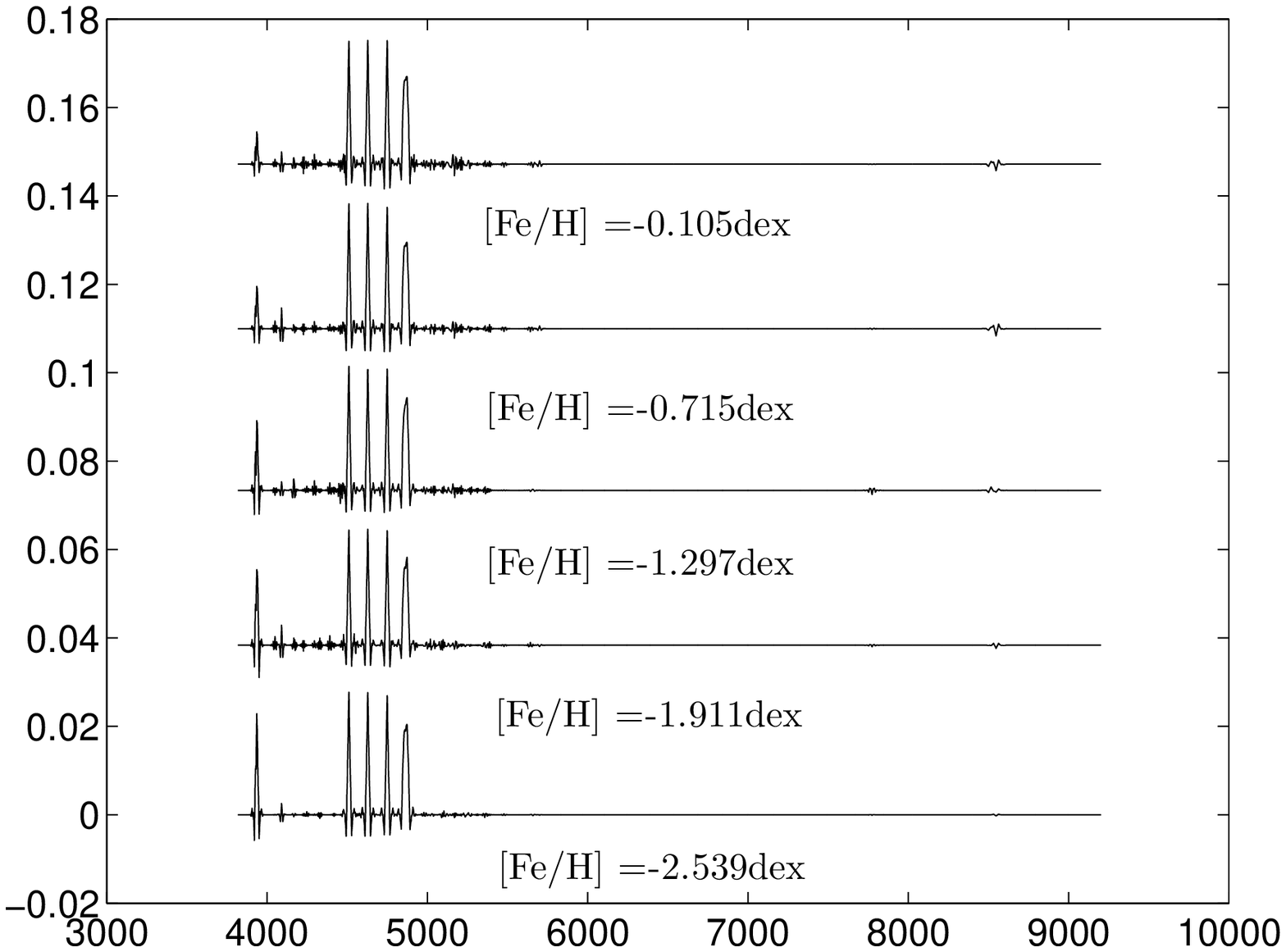}}
    \setlength{\abovecaptionskip}{-10pt}
  \caption{Visualization of the detected features (Section \ref{Sec:Spectrum_Decomposition:feature_selection}). Fig. \ref{Fig:feature:visualization:Teff2}, Fig. \ref{Fig:feature:visualization:logg2}, and Fig. \ref{Fig:feature:visualization:feh2} present the features in Table \ref{Tab:Detected_features} (a), Table \ref{Tab:Detected_features} (b), and Table \ref{Tab:Detected_features} (c) for the spectra drawn, respectively, in Fig. \ref{Fig:feature:visualization:Teff1}, Fig. \ref{Fig:feature:visualization:logg1}, and Fig. \ref{Fig:feature:visualization:feh1}. For example, the curve labeled with $T_{\texttt{eff}} = 9667K$ in Fig. \ref{Fig:feature:visualization:Teff2} is the visualization of the features in Table \ref{Tab:Detected_features} (a) for the spectrum labeled with $T_{\texttt{eff}} = 9667K$ in Fig. \ref{Fig:feature:visualization:Teff1}. }
  \label{Fig:feature:visualization} 
\end{figure*}

\begin{figure*}
  \centering
  \subfigure[Features for $T_{\texttt{eff}}$]{
    \label{Fig:feature:distribution:Teff} 
    \includegraphics[width =2in]{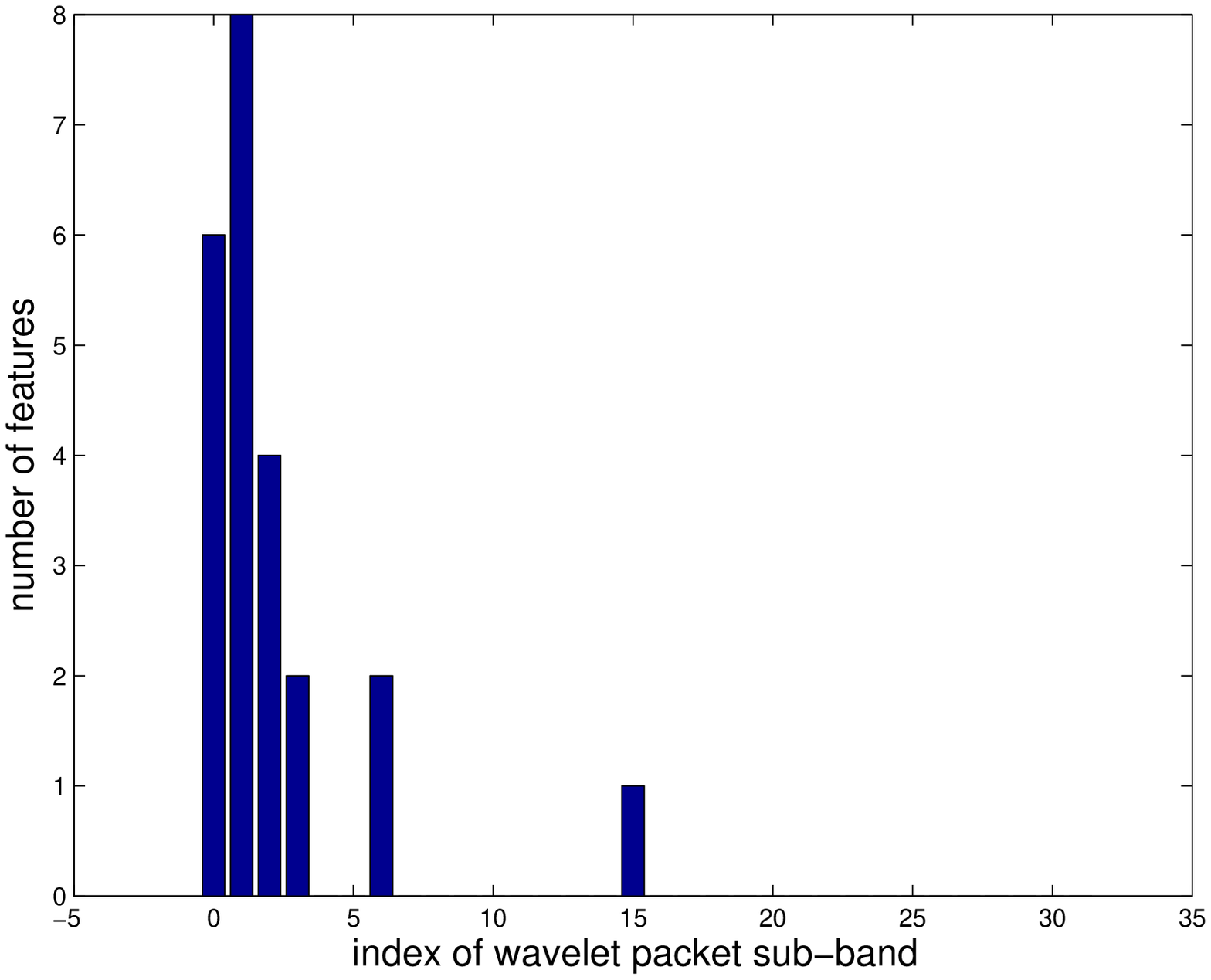}}
\hspace{-0.17in}
  \subfigure[Features for log~$g$]{
    \label{Fig:feature:distribution:logg} 
    \includegraphics[width =2in]{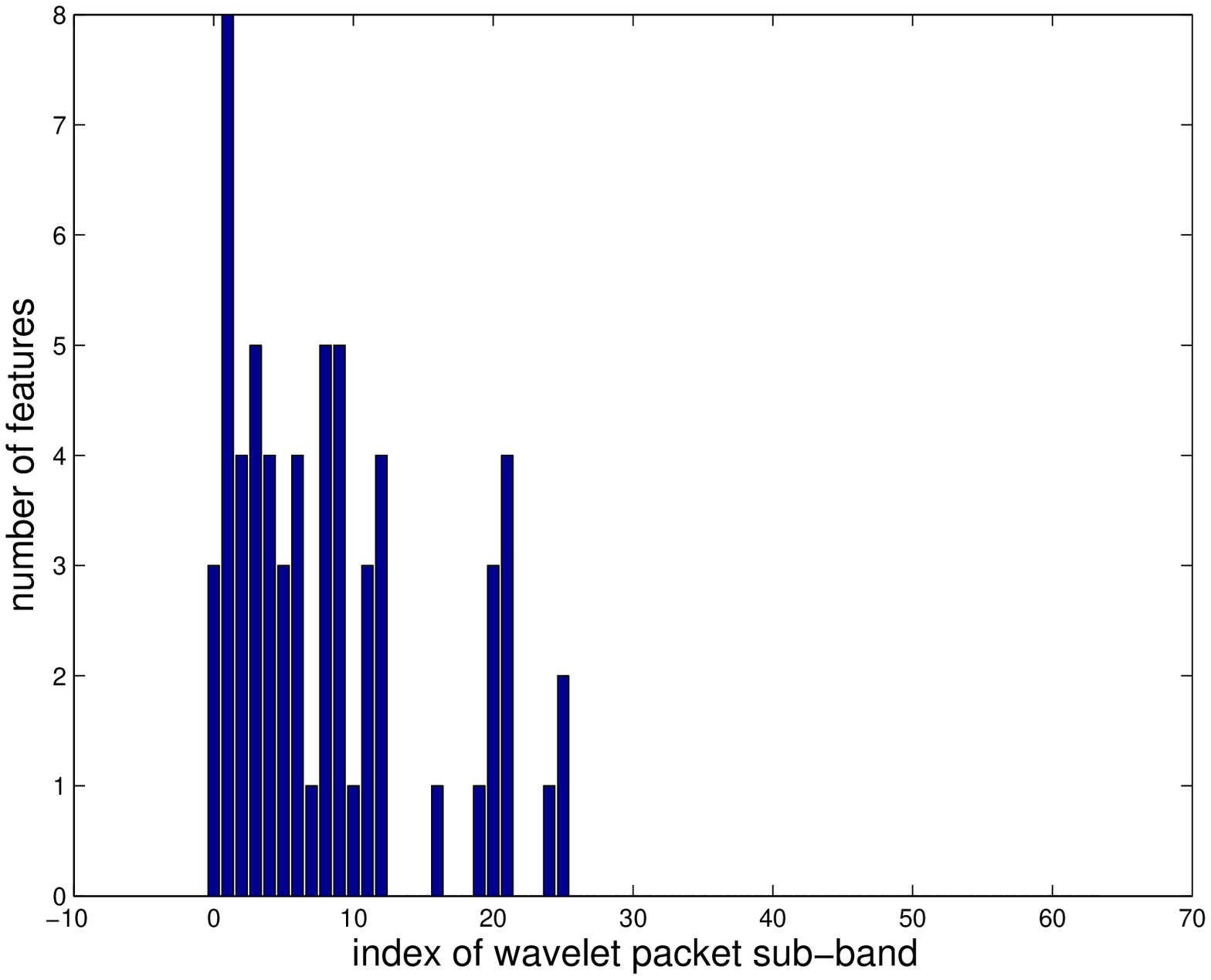}}
  \hspace{-0.17in}
  \subfigure[Features for $\texttt{[}$Fe/H$\texttt{]}$]{
    \label{Fig:feature:distribution:FeH} 
    \includegraphics[width =2in]{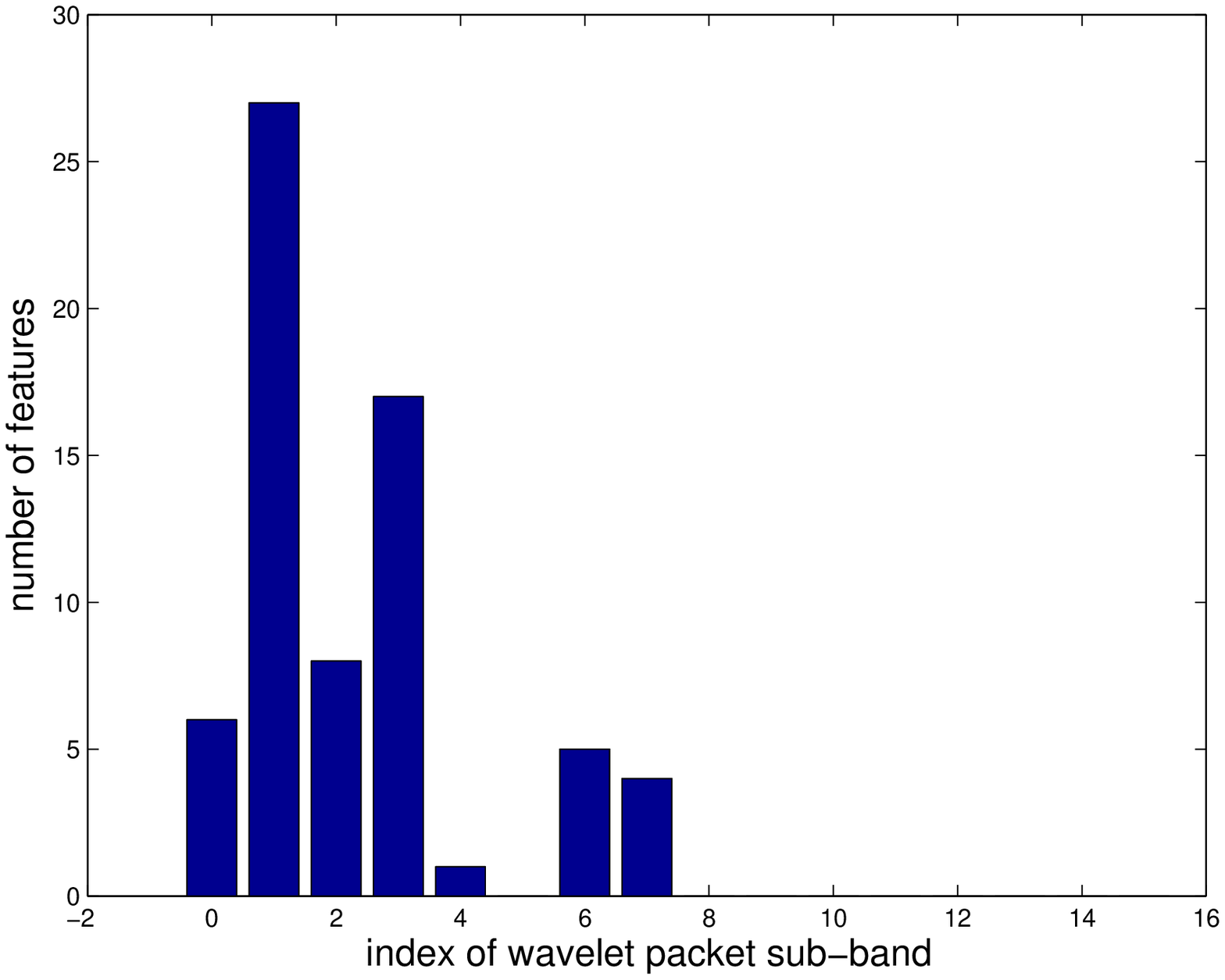}}
\hspace{-0.17in}
    \subfigure[Features for $T_{\texttt{eff}}$]{
    \label{Fig:feature:distribution:Teff_2Dimage} 
    \includegraphics[width =2in]{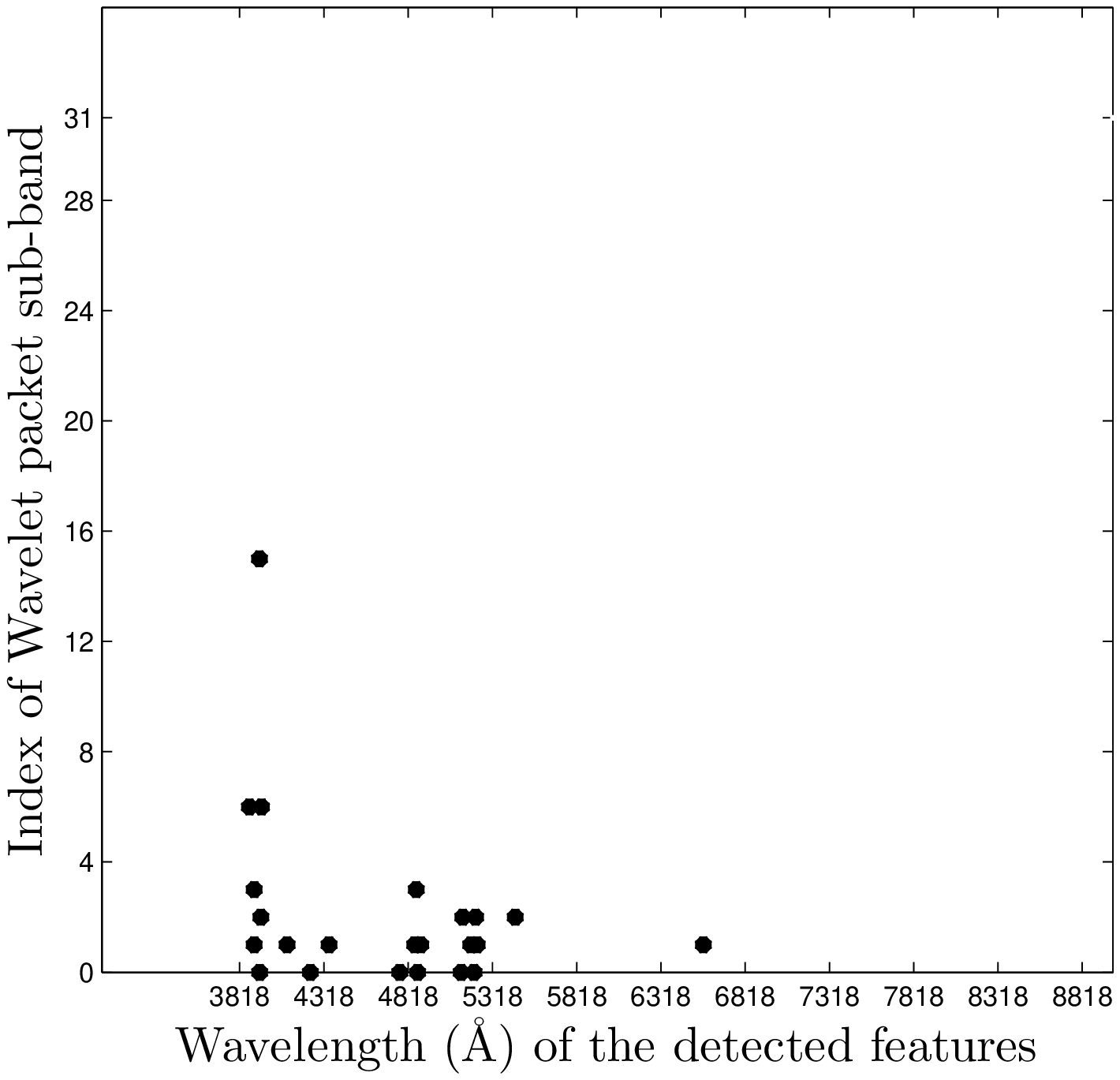}}
\hspace{-0.17in}
  \subfigure[Features for log~$g$]{
    \label{Fig:feature:distribution:logg_2Dimage} 
    \includegraphics[width =2in]{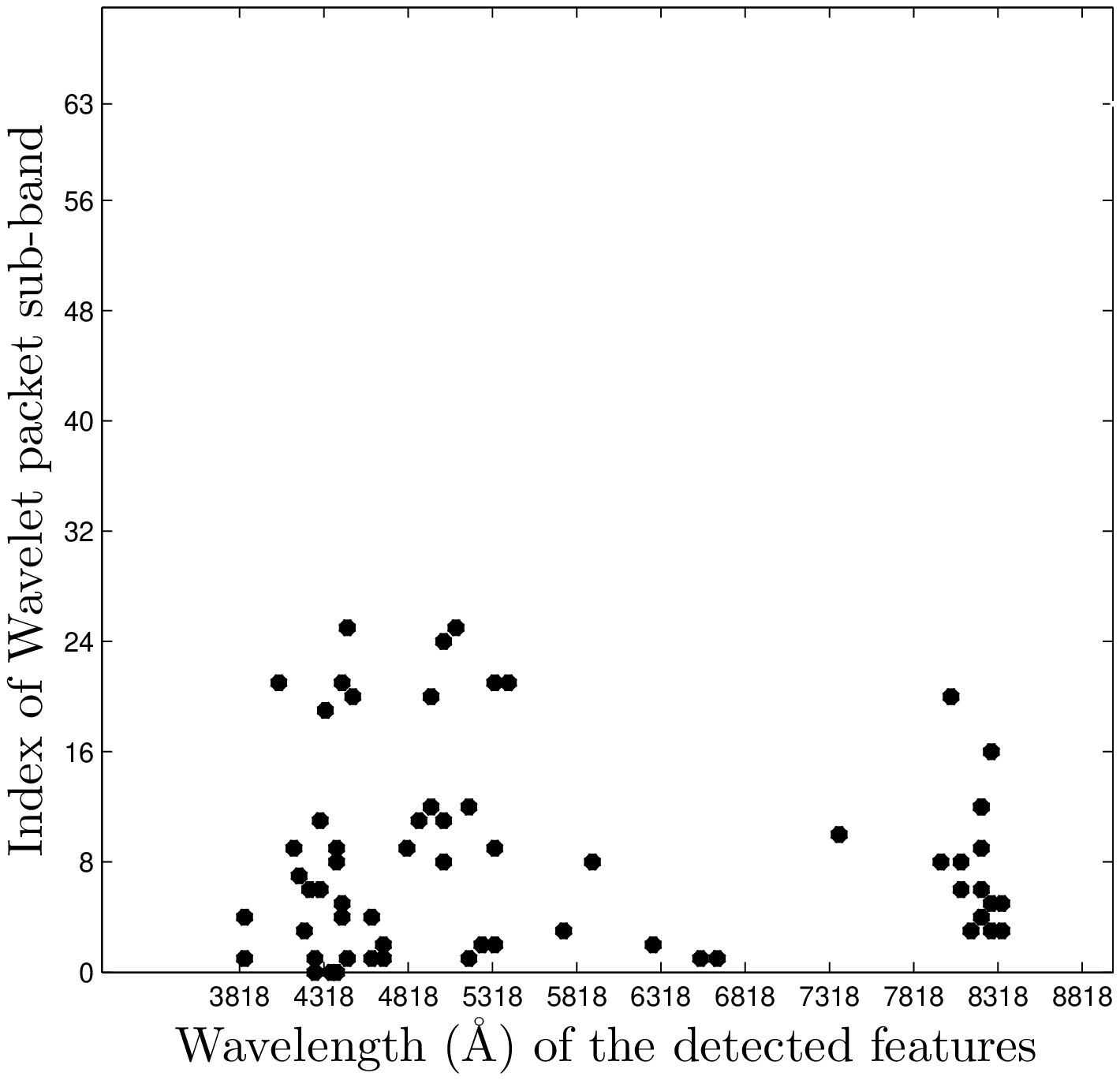}}
\hspace{-0.17in}
  \subfigure[Features for $\texttt{[}$Fe/H$\texttt{]}$]{
    \label{Fig:feature:distribution:FeH_2Dimage} 
    \includegraphics[width =2in]{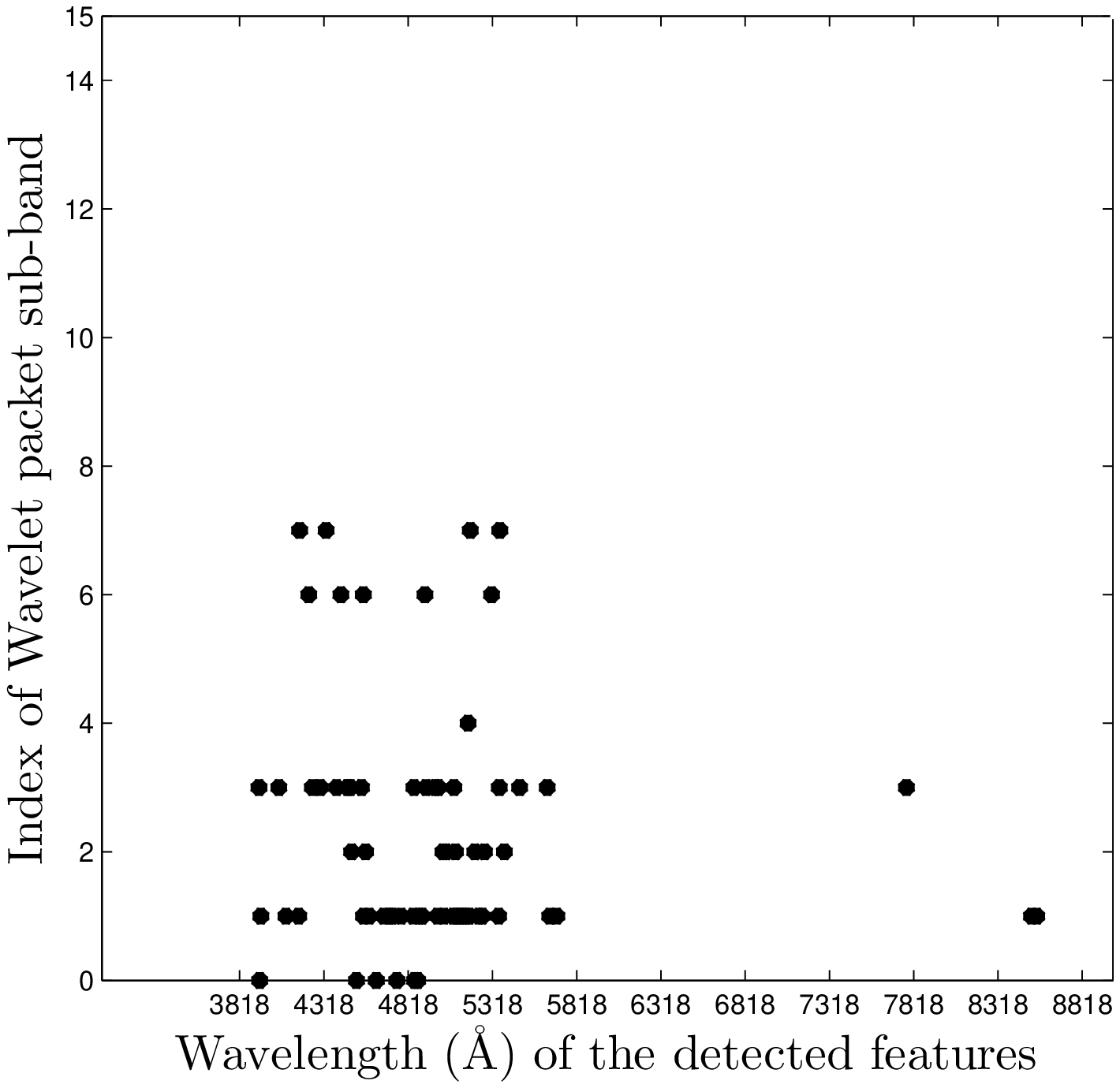}}
    \setlength{\abovecaptionskip}{-10pt}
  \caption{Distribution of the detected features in Table \ref{Tab:Detected_features}. (a) and (d) There are 120 wavelet packet components in each sub-band and more than 93.33\% of the components are redundancy and noise in each sub-band. (b) and (e) There are 60 wavelet packet components in each sub-band and more than 85\% of the components are redundancy and noise in each sub-band. (c) and (f) There are 239 wavelet packet components in each sub-band and more than 88.7\% of the components are redundancy and noise in each sub-band.(a) Features for $T_{\texttt{eff}}$; (b) Features for log~$g$; (c) Features for $\texttt{[}$Fe/H$\texttt{]}$; (d) Features for $T_{\texttt{eff}}$; (e) Features for log~$g$; and (f) Features for $\texttt{[}$Fe/H$\texttt{]}$; }
  \label{Fig:feature:distribution} 
\end{figure*}

\section{ESTIMATING THE  ATMOSPHERIC PARAMETERS}\label{Sec:estimation}
\subsection{Performance on SDSS Spectra}\label{Sec:estimation:SDSS}
Based on the detected features in Table \ref{Tab:Detected_features}, we can estimate the atmospheric parameters using the linear model defined in Equation (\ref{Equ:linear:model}), which can be learned from the training set (Section \ref{Sec:Data:SDSS}) by the OLS method in Equation (\ref{Sec:MSE:Obj}) and SVR method with a linear kernel (SVR$_l$) \citep{Journal:Schokopf:2002,Journal:Smola:2004}. The Performance of the test set (Section \ref{Sec:Data:SDSS}) is presented in Table \ref{Tab:effectiveness} (a). When using SVR$_l$, there is a regularization parameter $C$ that has to be preset \citep{Tec:Chang:2001,Journal:Schokopf:2002,Journal:Smola:2004}, and we optimized this parameter using the validation set(Section \ref{Sec:Data:SDSS}).

On the test set of 30,000 SDSS spectra, the formal MAE consistencies of the proposed scheme are 0.0062 dex for log$~T_\texttt{eff}$ (83 K for $T_\texttt{eff}$), 0.2345 dex for log$~g$, and 0.1564 dex for [Fe/H], where the MAE evaluation method is defined in Equation (\ref{Equ:MAE}). Therefore, the detected features provide excellent linear support for estimating atmospheric parameters $T_\texttt{eff}$, log$~g$ and [Fe/H].

In related work in the literature, the authors use various performance evaluation methods. In order to better compare with those sources, we have also made a performance evaluation of the proposed scheme based on measures of ME and SD (defined in equations (\ref{Equ:ME}) and (\ref{Equ:SD})): the results are presented in Table \ref{Tab:effectiveness} (a). Direct comparisons with published work are given in Section \ref{Sec:estimation:comparision}.

\begin{table*}\scriptsize
\setlength{\abovecaptionskip}{-100pt}
\setlength{\belowcaptionskip}{-100pt}
\centering
\caption{Performance of the proposed scheme}
\begin{tabular}{  c c c c c c}
\hline \hline
\multicolumn{6}{c}{(a) Performance on SDSS Test Set Consisting of 30,000 Stellar Spectra } \\
\hline
Estimation Method & Evaluation Method   &  log~$T_\texttt{eff}$ (dex)  &  $T_\texttt{eff}$ (K)  &       log$~g$ (dex)    &   [Fe/H](dex)  \\ \hline
\multirow{3}{2cm}{OLS}
                       &     MAE        &  0.0062                    &   82.94              &       0.2345         &   0.1564\\
                       &     ME         &  0.0002                    &   2.769              &      -0.0219         &   -0.0003\\
                       &     SD         &  0.0096                    &   135.9              &      0.3297         &   0.2196\\ \hline
\multirow{3}{2cm}{SVR$_l$}
                       &     MAE        &  0.0060                    &   80.67              &       0.2225         &   0.1545\\
                       &     ME         &  0.0002                    &   4.783               &      -0.0762        &    -0.0012\\
                       &     SD         &  0.0096                    &   136.6               &      0.3298         &   0.2177\\

\hline \hline
\multicolumn{6}{c}{(b) Performance on Synthetic Test Set Consisting of 8500 Spectra } \\
\hline
Estimation Method & Evaluation Method   &  log~$T_\texttt{eff}$ (dex)  &  $T_\texttt{eff}$ (K)  &       log$~g$ (dex)    &   [Fe/H](dex)  \\ \hline
\multirow{3}{2cm}{OLS}
                       &     MAE        &  0.0022                    &   31.69              &   0.0337             &   0.0268\\
                       &     ME         &  0.0003                    &   2.823              &   -0.0004            &   0.0049\\
                       &     SD         &  0.0029                    &   41.45              &   0.0687             &   0.0371\\ \hline

\multirow{3}{2cm}{SVR$_l$}
                       &     MAE        &  0.0031                    &   43.74              &       0.0611         &   0.0359\\
                       &     ME         &  -0.0001                   &   -2.886             &      0.0025          &   0.0024\\
                       &     SD         &  0.0040                    &   58.74              &      0.0966          &   0.0514\\ \hline
\hline
\\
\multicolumn{6}{p{13cm}}{Note. The number of extracted features is 23 for $T_\texttt{eff}$, 62 for log$~g$, and 68 for [Fe/H]. OLS (Ordinary Least Squares): linear least squares regression, SVR$_l$: Support Vector machine Regression with a linear kernel.}
\end{tabular}\label{Tab:effectiveness}
\end{table*}

\subsection{Performance on Synthetic Spectra with Ground-truth}\label{Sec:estimation:synthetic}
The proposed scheme is also evaluated on synthetic spectra built from theoretical parameters. The synthetic data set is described in Section \ref{Sec:Data:Synthetic}. This experiment shares the same parameters as the experiment using SDSS data to detect features with LASSO(LARS)$_{\texttt{bs}}$ -- m is 100, $k_0$ are 23 for $T_\texttt{eff}$, 62 for log$~g$, and 68 for [Fe/H].

For the test set of 8500 synthetic spectra, the MAE accuracies when the OLS estimation is used are 0.0022 dex for log$~T_\texttt{eff}$ (32 K for $T_\texttt{eff}$), 0.0337 dex for log$~g$, and 0.0268 dex for [Fe/H]. More results are presented in Table \ref{Tab:effectiveness} (b).

 When experimenting with real spectra, results usually are influenced by noise and calibration defects. Therefore, SVR$_l$ are slightly more accurate than OLS because it incorporates a regularization technique (Table \ref{Tab:effectiveness} (a)). For synthetic spectra in which no external disturbances occur, OLS are more accurate than SVR$_l$ (Table \ref{Tab:effectiveness} (b)).

\subsection{Comparison with Previous Works} \label{Sec:estimation:comparision}
The proposed scheme is tested on both real spectra from SDSS and synthetic spectra derived from Kurucz's NEWODF models \citep{Journal:Castelli:2003}. Using large spectral samples from SDSS and synthetic stellar models, several authors have attempted to obtain accurate estimates of atmospheric parameters along similar scenarios. These works can be classified into two groups based on the estimation methods: linear schemes and nonlinear schemes.
\begin{enumerate}
\item \emph{Nonlinear methods}:~~~~
\citet{Journal:Fiorentin:2007} investigated the stellar parameter estimation problem based on Principal Component Analysis (PCA) and nonlinear artificial neural networks (ANN) and obtained MAE accuracies 0.0126 dex for log$~T_\texttt{eff}$, 0.3644 dex for log$~g$, and 0.1949 dex for [Fe/H] in a test set of 19,000 stellar spectra from SDSS. \citet{Journal:Jofre:2010} applied a nonlinear MA$\chi$ method to a sample set of 17,274 spectra of metal-poor dwarf stars from SDSS/SEGUE and estimated the effective temperature, log$~g$, and the metallicity with respective average accuracies of 130 K (ME), 0.5 dex (ME), and 0.24 dex (ME). \citet{Journal:Li:2014} used a LASSO scheme coupled with nonlinear SVR$_G$ (Support Vector Regression with a Gaussian kernel) and reached MAE accuracies of 0.0075 dex for log$~T_\texttt{eff}$
(101.6 K for $~T_\texttt{eff}$ ), 0.1896 dex for log$~g$, and 0.1821 for [Fe/H].
\item \emph{Linear methods}:~~~~\citet{Journal:Tan:2013} used a Lick line index of SDSS spectra and a linear regression method: the SD accuracies are 196.5 K for $T_\texttt{eff}$, 0.596 dex for log$~g$, and 0.466 dex for [Fe/H]. \citet{Journal:Li:2014} also studied the physical parameter estimation problem using LASSO and the SVR$_l$ with MAE accuracies 0.0342 dex for log$~T_\texttt{eff}$, 0.2534 dex for log$~g$, and 0.3235 for [Fe/H].
\end{enumerate}

Finally, \citet{Journal:Fiorentin:2007}, using a test set of 908 synthetic spectra calculated from  Kurucz's NEWODF models \citep{Journal:Castelli:2003}, applied PCA and nonlinear ANN, and obtained test accuracies of 0.0030 dex for log$~T_\texttt{eff}$, 0.0245 dex for log$~g$, and 0.0269 dex for [Fe/H] \citep[Table 1,][]{Journal:Fiorentin:2007}.

The literature results are summarized in Table \ref{Tab:effectiveness:comparision}. It can be seen that the scheme proposed in the present paper provides excellent performance when estimating stellar atmospheric parameters.

\renewcommand{\multirowsetup}{\centering}
\begin{table*}\scriptsize
\setlength{\abovecaptionskip}{-100pt}
\setlength{\belowcaptionskip}{-100pt}
\centering
\caption{Comparing the Proposed Scheme with Previous Works in Similar Scenarios}
\begin{tabular}{ l c c c c c c}
\hline \hline
\multicolumn{7}{c}{(a) Comparison with SDSS Data Set. } \\
\hline
Estimation Method & Evaluation Method   &  log~$T_\texttt{eff}$ (dex)  &  $T_\texttt{eff}$ (K)  &       log$~g$ (dex)    &   [Fe/H](dex) & Size of Test Set \\ \hline
\multirow{3}{1cm}{Linear:OLS}
                       &     MAE        &  0.0062                    &   82.94              &       0.2345         &   0.1564 & \multirow{3}{1cm}{30,000}\\
                       &     ME         &  0.0002                    &   2.769              &      -0.0219         &   -0.0003\\
                       &     SD         &  0.0096                    &   135.9              &      0.3297         &   0.2196\\ \hline
Nonlinear:ANN [1] & MAE & 0.0126                       &  -                      &      0.3644           & 0.1949 & 19,000\\
Nonlinear:MA$\chi$ [2] & ME &  -                           &130                      &      0.5              & 0.24& 17,274\\
Nonlinear:SVR$_G$[3] &MAE   &0.007458           & 101.610                 & 0.189557              & 0.182060 & 20,000\\
\hline
Linear:OLS [4] &SD             & -                &  196.473              & 0.596                  & 0.466 & 9048\\
Linear:SVR$_l$ [3]&  MAE & 0.034152    & -                      & 0.253363               & 0.323512 & 20,000\\
\hline \hline
\multicolumn{7}{c}{(b) Comparison with Synthetic Data Set Derived from Kurucz's NEWODF Models \citep{Journal:Castelli:2003} } \\
\hline
Estimation Method & Evaluation Method   &  log~$T_\texttt{eff}$ (dex)  &  $T_\texttt{eff}$ (K)  &       log$~g$ (dex)    &   [Fe/H](dex) & Size of Test Set  \\ \hline
Linear:OLS        &     MAE        &  0.0022                    &   31.70              &   0.0337             &   0.0268 & 30,000\\
Nonlinear:ANN [1] & MAE & 0.0030             &   -                    & 0.0245                 & 0.0269 & 19,000\\
\hline
\\
\multicolumn{7}{p{15cm}}{Note. OLS (Ordinary Least Squares): linear least squares regression, SVR$_l$: Support Vector machine Regression with a linear kernel, SVR$_G$: Support Vector machine Regression with a Gaussian kernel, ANN: Artificial neural networks, MA$\chi$: MAssive compression of $\chi^2$. [1]:\citet{Journal:Fiorentin:2007}, [2]:\citet{Journal:Jofre:2010}, [3]:\citet{Journal:Li:2014}, [4]:\citet{Journal:Tan:2013}.}
\end{tabular}\label{Tab:effectiveness:comparision}
\end{table*}

\section{MORE TECHNICAL DISCUSSIONS}\label{Sec:More:Technical}

\subsection{Configuration for WPD}\label{Sec:More:Technical:Config_wavelet_Decom}
We will now investigate the influence of the selection of wavelet basis functions and WPD level on atmospheric parameter estimates using the 10,000 SDSS spectra of the validation set. The considered basis functions include Biorthogonal basis (bior), Coiflets (coif), Daubechies basis (db), Haar (haar), ReverseBior (rbio), and Symlets (sym) \citep{Book:Mallat:2009}.

Consider $S_{wb}$ as a set of basis functions\\
$$
S_{wb} = \{bior, coif, db, haar, rbio, sym\},
$$
$S_l$ as a set of options for the WPD level
$$
S_{l} = \{3, 4, 5, 6, 7\},
$$
and $S_k$ as a set of options for $k_0$ in the proposed algorithm LASSO(LARS)$_{\texttt{bs}}$ in Section \ref{Sec:sub:Model_Further}.

The configuration optimization problem can be formulated as the search for
\begin{equation}\label{Equ:config:obj}
   \min\limits_{\texttt{wb} \in S_{\texttt{wb}}, \texttt{level} \in S_l, k_0 \in S_k }{MAE(\texttt{wb}, \texttt{level}, k_0 ,\texttt{ap})},
\end{equation}
where ap =$T_\texttt{eff}$, log$~g$ or [Fe/H], with MAEs being the evaluation of the predicted error for a specific configuration of wb, level, $k_0$, and ap .

\begin{figure*}
  \centering
  \subfigure[For $T_{\texttt{eff}}$]{
    \label{Fig:selection:waveletlevel:Teff} 
    \includegraphics[width =2in]{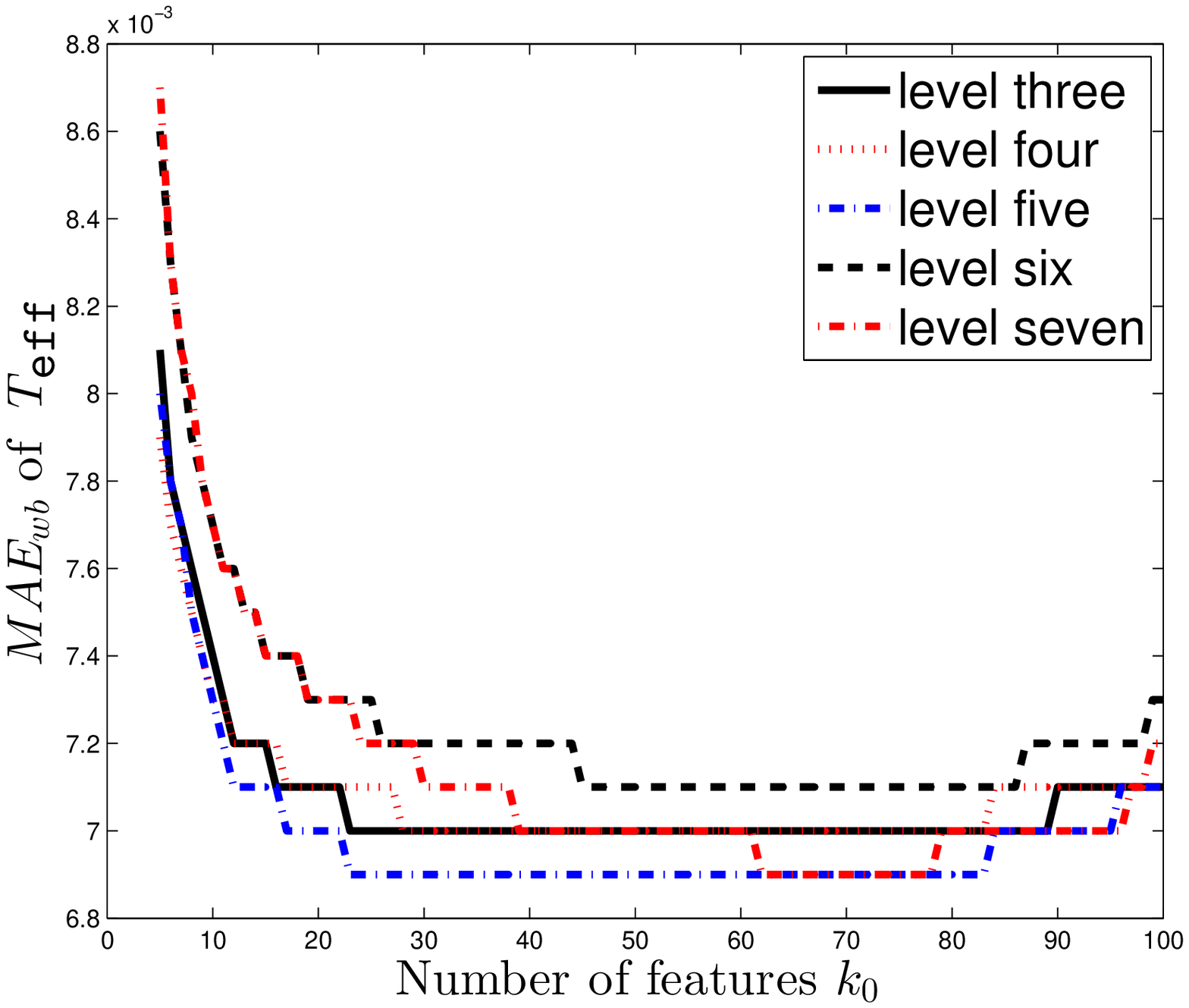}}
\hspace{-0.17in}
  \subfigure[For log~$g$]{
    \label{Fig:selection:waveletlevel:logg} 
    \includegraphics[width =2in]{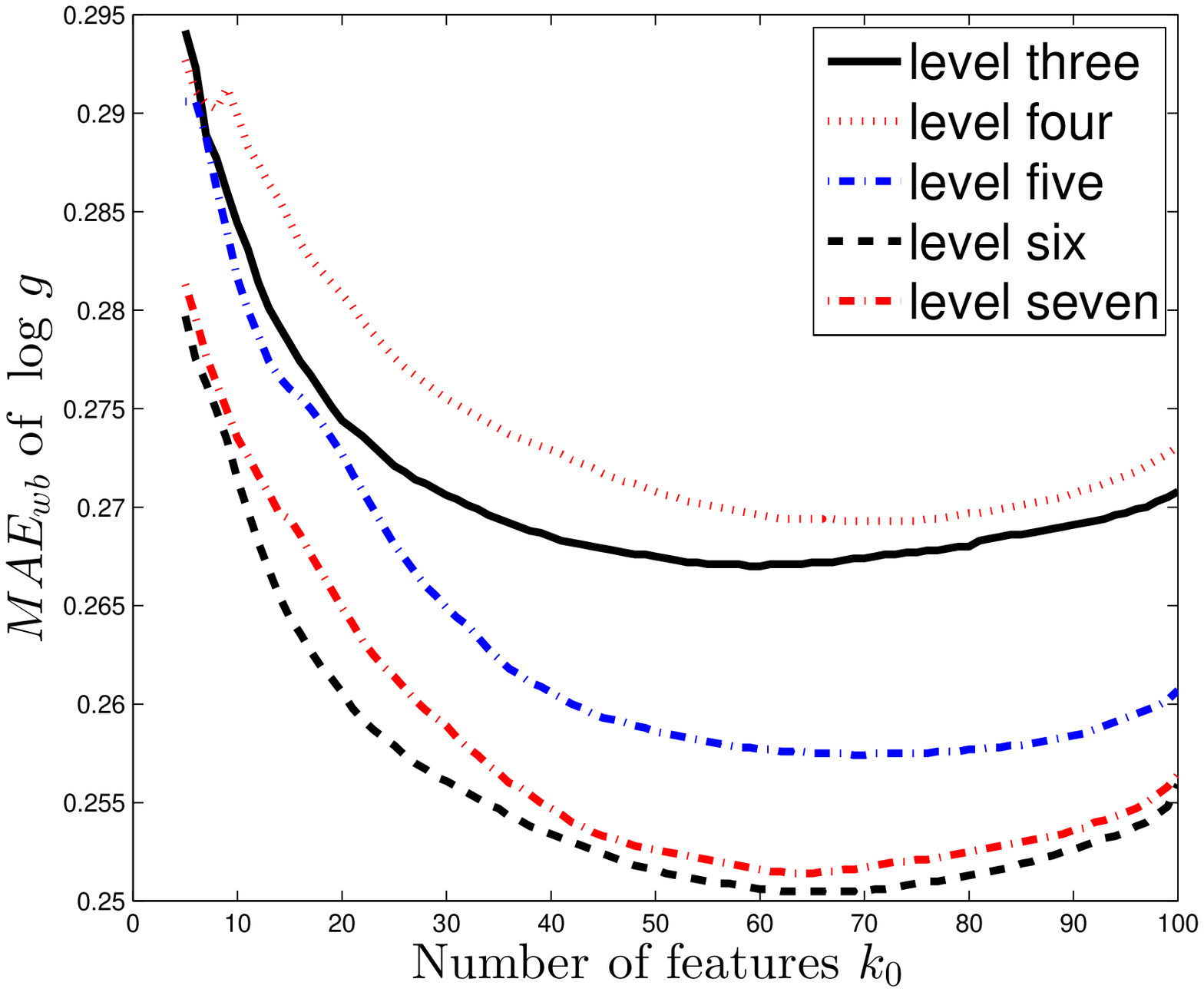}}
  \hspace{-0.17in}
  \subfigure[For $\texttt{[}$Fe/H$\texttt{]}$]{
    \label{Fig:selection:waveletlevel:FeH} 
    \includegraphics[width =2in]{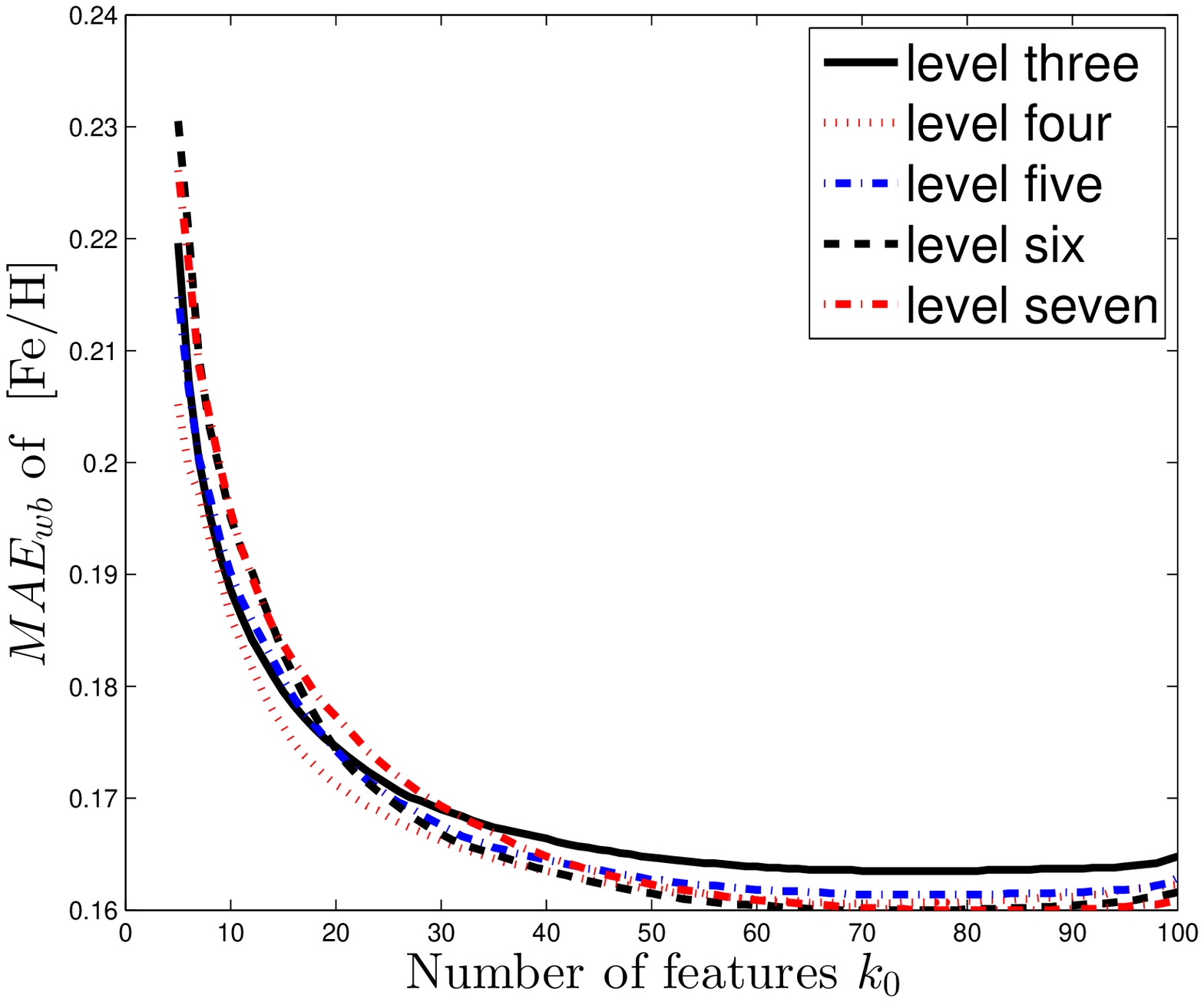}}
  \hspace{-0.17in}
  \subfigure[For $T_{\texttt{eff}}$]{
    \label{Fig:selection:waveletbase:Teff} 
    \includegraphics[width =2in]{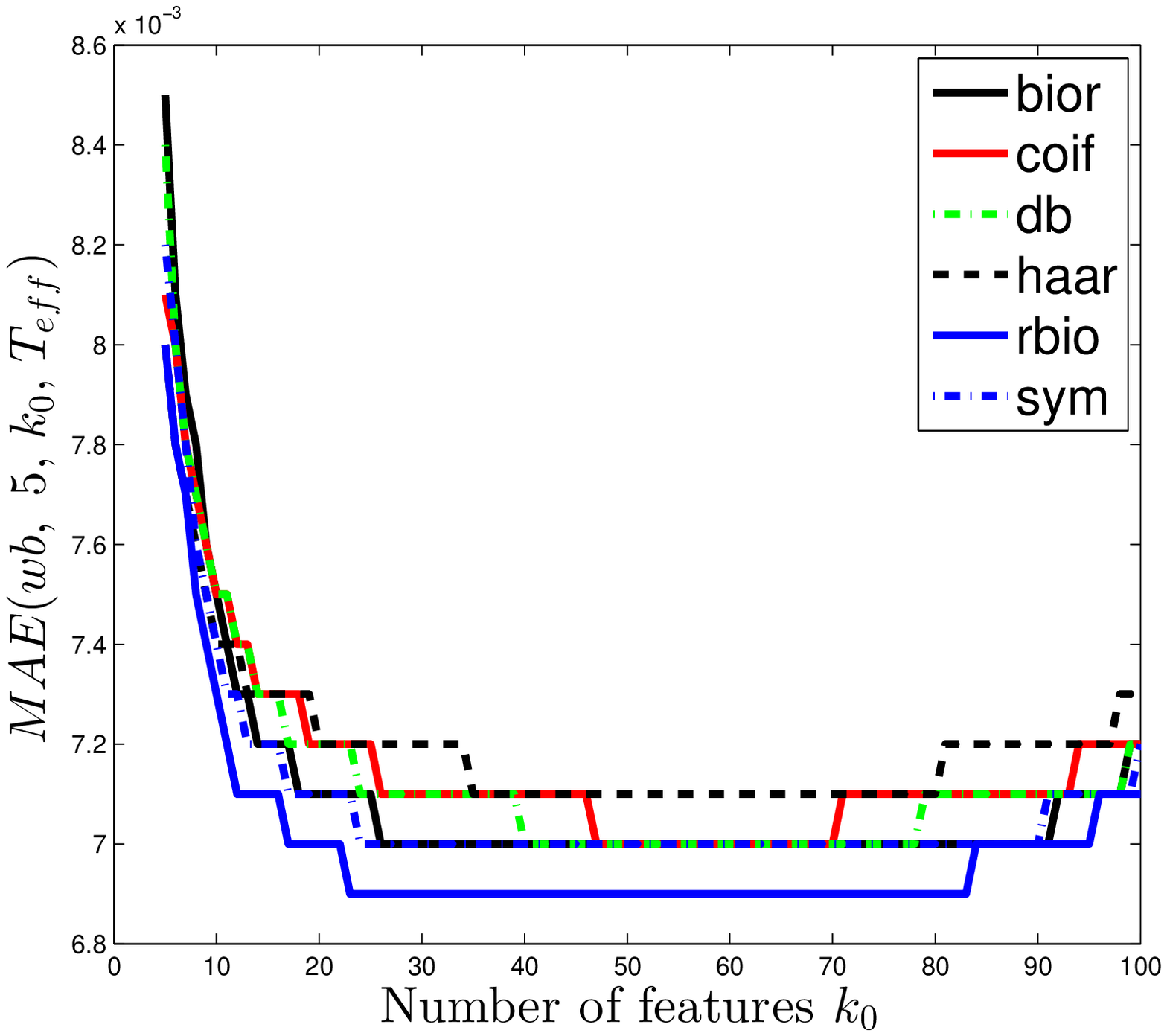}}
\hspace{-0.17in}
  \subfigure[For log~$g$]{
    \label{Fig:selection:waveletbase:logg} 
    \includegraphics[width =2in]{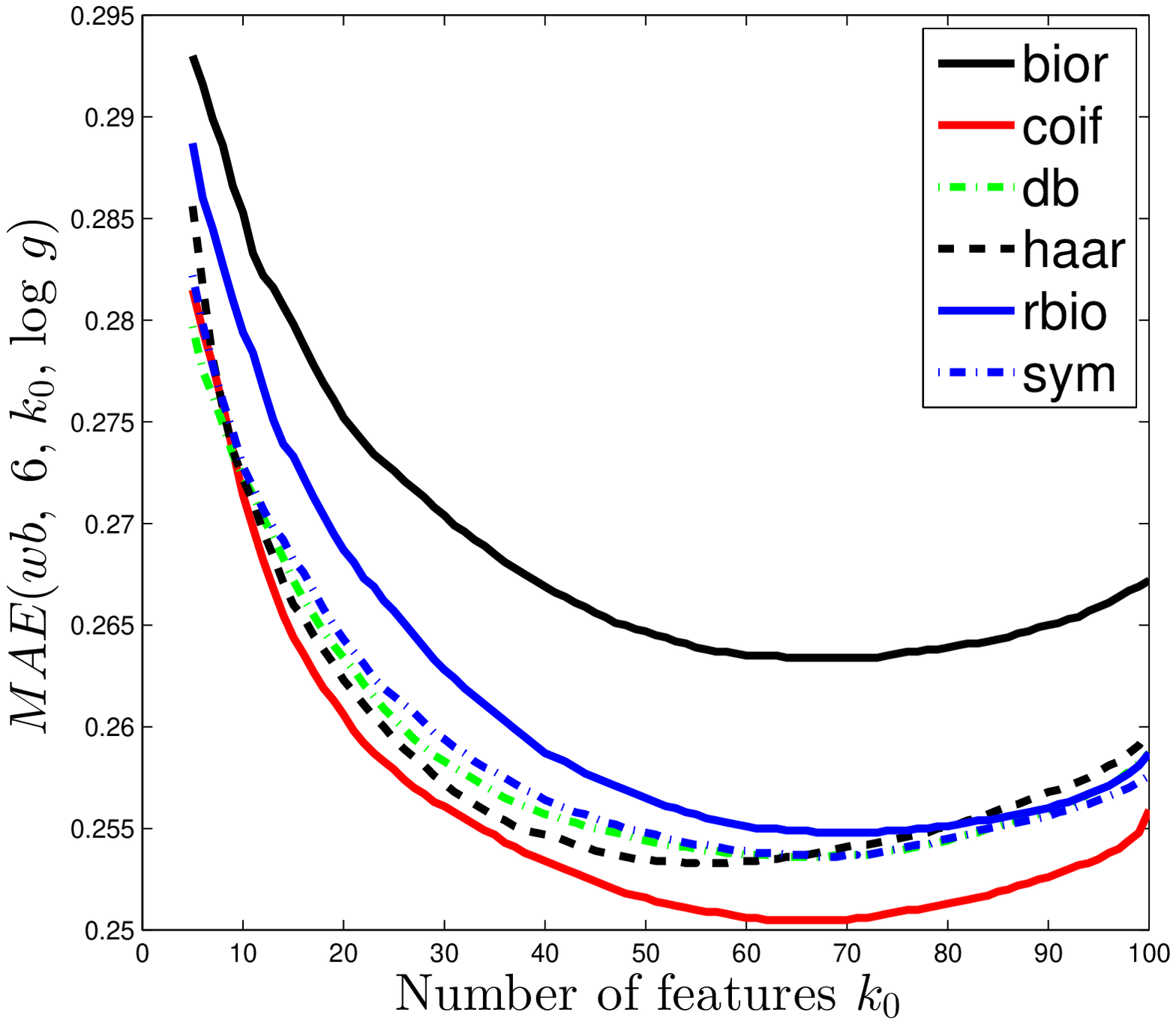}}
  \hspace{-0.17in}
  \subfigure[For $\texttt{[}$Fe/H$\texttt{]}$]{
    \label{Fig:selection:waveletbase:FeH} 
    \includegraphics[width =2in]{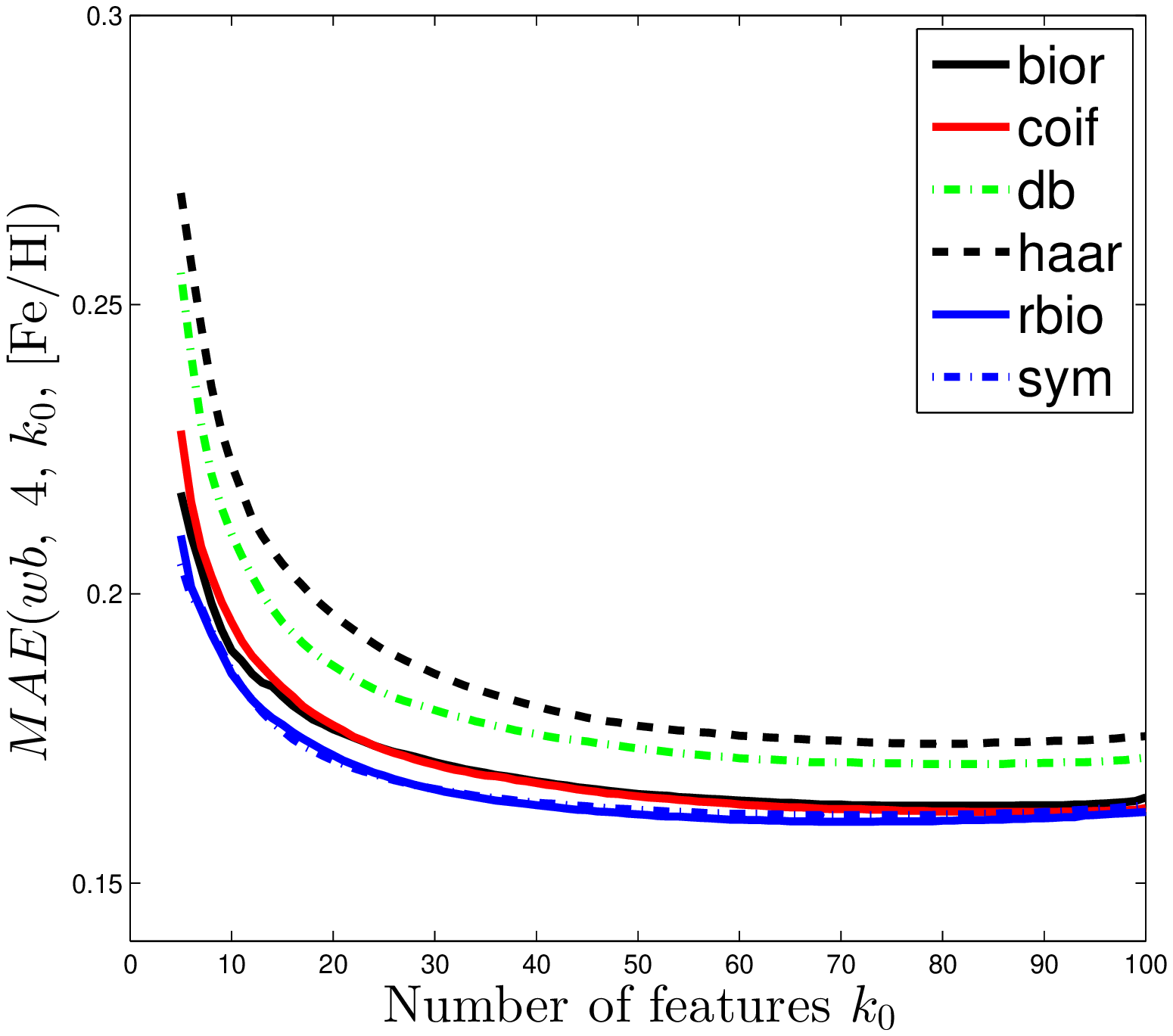}}
    \setlength{\abovecaptionskip}{-10pt}
  \caption{Optimize the configuration for wavelet packet decomposition. (a),(b), and (c) Selection of Wavelet packet decomposition level: the optimal decomposition levels are 5 for $T_\texttt{eff}$, 6 for log$~g$, and 4 for [Fe/H]. (d), (e), and (f) selection of wavelet basis function: the optimal basis functions and feature numbers are, respectively, rbio and 23 for $T_\texttt{eff}$, coif and 62 for log$~g$, and rbio and 68 for [Fe/H] based on the criterion in Equation (\ref{Equ:config:obj}). These experiments are conducted on the 10,000 SDSS spectra of the validation set (Section \ref{Sec:Data:SDSS}). (a) For $T_{\texttt{eff}}$, (b) for log~$g$, (c) for $\texttt{[}$Fe/H$\texttt{]}$, (d) for $T_{\texttt{eff}}$, (e) for log~$g$, (f) for $\texttt{[}$Fe/H$\texttt{]}$.}
  \label{Fig:selection:waveletpacket} 
\end{figure*}

We initially select $m =100$ features using the LASSO(LARS)$_{\texttt{bs}}$ scheme and let
$$
  S_k = \{ 100, 99, 98, \cdots, 5 \}.
$$
To obtain the optimal decomposition level, let
\begin{equation}\label{Equ:min_wb}
\begin{split}
            & MAE_{wb}(level, k_0 ,ap) \\
\triangleq  & \min\limits_{wb \in S_{wb}}{MAE(wb, level, k_0 ,ap)}.
\end{split}
\end{equation}
The relationship between $MAE_{wb}$ and $k_0$ are investigated on SDSS spectra for every combination of $level \in S_l$ and $ap = \{T_\texttt{eff}, log~g, [\texttt{Fe/H}]\}$, and the experimental results are presented in Fig. \ref{Fig:selection:waveletlevel:Teff}, Fig. \ref{Fig:selection:waveletlevel:logg} and Fig. \ref{Fig:selection:waveletlevel:FeH}. The optimal WPD levels appear to be 5 for $T_\texttt{eff}$, 6 for log$~g$, and 4 for [Fe/H] based on the criterion defined in Equation (\ref{Equ:config:obj}).

Once the optimal decomposition level has been found, the performances of various basis functions are investigated and the associated optimal number of features can be derived. The experimental results are presented in Fig. \ref{Fig:selection:waveletbase:Teff}, Fig. \ref{Fig:selection:waveletbase:logg} and Fig. \ref{Fig:selection:waveletbase:FeH}. Based on the criterion defined in Equation (\ref{Equ:config:obj}), we find that the optimal basis functions and feature numbers are, respectively, rbio and 23 for $T_\texttt{eff}$, coif and 62 for log$~g$, and rbio and 68 for [Fe/H].

\subsection{Sufficiency and Compactness}\label{Sec:More:Technical:Sufficiency_compactness}
We now explore the sufficiency of the set of LASSO(LARS)$_{\texttt{bs}}$ detected features as defined in Table \ref{Tab:Detected_features}; that is, we study whether the accuracy of the atmospheric parameter estimation can be significantly improved by appending some additional components of the WPD.

To do this, we conduct six experiments by appending the components of WPD having the lowest frequency or the highest frequency in the LASSO(LARS)$_{\texttt{bs}}$ feature set. The corresponding results are presented in rows (3) and (4) of Table \ref{Tab:effectiveness:noise_redundancy}. For convenience, the performance of the LASSO(LARS)$_{\texttt{bs}}$ features is repeated in row (1) of Table \ref{Tab:effectiveness:noise_redundancy}.

It appears that the performance gain is trivial after adding more components to the LASSO(LARS)$_{\texttt{bs}}$ features. The WP components with the lowest frequency are the traditional choice of spectral features for estimating atmospheric parameters \citep{Journal:Lu:2013}. If we add them to $\{L_i\}$, the amount of features will increases from 62 to 144 (increase 132.26\%), but the MAEs can only decrease 0.0101 (4.3\%). Adding them to $\{F_i\}$, the amount of features increases 354.41\% and the MAE only decrease 6.65\%. On the other hand, if we add the components with the highest frequency to the features in Table \ref{Tab:Detected_features}, it is shown that the performance decreases (row (4) in Table \ref{Tab:effectiveness:noise_redundancy}). Therefore, we conclude that the detected features in Table \ref{Tab:Detected_features} are quite sufficient.

Suppose $S1$ and $S2$ are two sets of features, then $||S1||$ and $||S2||$ represent the number of features in $S1$ and $S2$ respectively. If $||S1|| <||S2||$, then we will say that $S1$ is more compact than $S2$.

We also investigate the performance of the traditional choice of the components with the lowest frequency, and the results are presented in row (2) in Table \ref{Tab:effectiveness:noise_redundancy}. It is shown that the accuracy and compactness of the features in Table \ref{Tab:Detected_features} are all better than those of the components with the lowest frequency.

\begin{table*}\scriptsize
\setlength{\abovecaptionskip}{-100pt}
\setlength{\belowcaptionskip}{-100pt}
\centering
\caption{Sufficiency and Compactness of the Detected Features Identified in Table \ref{Tab:Detected_features}}
\begin{tabular}{c|c|c|c|c|c|c}
  \hline \hline
label & \multicolumn{2}{c|}{log~$T_\texttt{eff}$}  &  \multicolumn{2}{c|}{log~$g$}  & \multicolumn{2}{c}{[Fe/H]}   \\
\hline
 (1) & $\{T_{i}\}$:23                  &  0.0062 &   $\{L_{i}\}$:62                 &  0.2351    &   $\{F_{i}\}$:68                 &  0.1564  \\  \hline
 (2) &  WP(rbio,5,0):128                &  0.0068 &   WP(coif,6,0):82                &  0.2482    &   WP(rbio,4,0):247               &  0.1573   \\ \hline
 (3) &  WP(rbio,5,0)+$\{T_{i}\}$:145    &  0.0062 &   WP(coif,6,0)+$\{L_{i}\}$:144   &  0.2250    &   WP(rbio,4,0) +$\{F_{i}\}$:309 &  0.1460  \\  \hline
 (4) &  WP(rbio,5,31)+$\{T_{i}\}$:151   &  0.0063 &   WP(coif,6,63)+$\{L_{i}\}$:144  &  0.2364    &   WP(rbio,4,15)+$\{F_{i}\}$:315 &  0.1608  \\  \hline
 \\
 \multicolumn{7}{p{15cm}}{Note. In these experiments, the atmospheric parameters are estimated by OLS method and the performance is evaluated by MAEs. WP($wp$, $i$, $j$): Decompose a spectrum by wavelet packet transform based on basis $wp$, and take the components in the $j$th sub-band at level $i$ as features. $\{T_{i}\}$, $\{L_{i}\}$, $\{F_{i}\}$ represent the features $\{T_{i}, i = 1,\cdots, 23 \}$, $\{L_{i}, i = 1,\cdots, 62\}$, $\{F_{i}, , i = 1,\cdots, 68 \}$ in Table \ref{Tab:Detected_features}, respectively. The number behind `:' represent the number of selected features in a specific experiment.}
\end{tabular}\label{Tab:effectiveness:noise_redundancy}
\end{table*}

\subsection{Redundancy - Positive or Negative?}
Redundancy is the duplication of some components in a system. In previous sections, the removal of redundant features is discussed as positive for performance improvement, but should not redundancy be useful in the presence of noise?

\emph{Potential advantages of redundancy~~~~}In theory, redundancy should help to remove noise or at least reduce the negative effects of noise. The usefulness depends on the relationship between components and their duplicates. Unfortunately, it is difficult to uncover these relationships. Multiple independent components are usually present, and the duplicates of these different components are often mixed up and hard to identify, which makes an effective use of redundancy very difficult to implement.

\emph{Potential disadvantages~~~~}The existence of redundancy, in addition to an increase in computational burden, usually destroys or reduces the quality of investigations based on computer algorithms. The learning process of computer algorithms can be regarded as some kind of vote assessment. In applications, the existing components usually differ from each other in the limit of redundancy, but usually the amount of redundancy for a specific component remains unknown. Thus, multiple components in data invisibly assume different number of votes, which usually results in erroneous evaluation results and reduces the quality of learning\footnote{In theory, every component should ideally only take one effective vote for fairness in a condition.}.

\section{CONCLUSION}\label{Sec:Conclusion}

We propose a scheme LASSO(LARS)$_{\texttt{bs}}$ to extract linearly supporting (LSU) features from stellar spectra to estimate the atmospheric parameters $T_\texttt{eff}$, log$~g$, and [Fe/H]. `Linearly supporting' means that the atmospheric parameters can be accurately estimated from the extracted features using a linear model. One prominent characteristic of the proposed scheme is the ability to directly evaluate the contribution of the detected features to the estimate of the atmospheric parameters (Table \ref{Tab:coefficients}) and to trace back the physical interpretation of the features (Section \ref{Sec:Spectrum_Decomposition:characteristic}).

The basic idea of this work is that the effectiveness of a data component is sensitive to both wavelength and frequency. Therefore, we decompose the stellar spectra using WPs before detecting features. It is shown that at most 1.72\% of the data components are necessary features for estimating atmospheric parameters (Table \ref{Tab:Detected_features}), and LASSO(LARS)$_{\texttt{bs}}$ can effectively delete the redundancy and noise (Fig. \ref{Fig:feature:distribution}). The detected features are sparse.

Due to the time--frequency localization of WPD, we can derive the wavelength of the detected features (in the spectral space; Fig. \ref{Fig:feature:distribution:Teff_2Dimage}, Fig. \ref{Fig:feature:distribution:logg_2Dimage}, Fig. \ref{Fig:feature:distribution:FeH_2Dimage} and Table \ref{Tab:Detected_features}). The feature wavelength position helps us to identify the selected features with specific spectral lines, which leads to an understanding of the physical significance of the detected features (Section \ref{Sec:Spectrum_Decomposition:characteristic}).

The accuracies/consistencies of the proposed scheme LASSO(LARS)$_{\texttt{bs}}$ + OLS with respect to the pre-estimation by SSPP of SDSS for real spectra and with respect to exact input atmospheric parameters in stellar models are evaluated through three statistical indicators and compared with previous similar works in the literature. The proposed scheme is shown to provide excellent performances, both on real (noisy) spectra and on synthetic stellar models, and therefore, the detected features provided excellent linear support when estimating the atmospheric parameters $T_\texttt{eff}$, log$~g$ and [Fe/H].

\acknowledgments
The authors would like to thank the reviewer and editor for their instructive comments, and extend their thanks to
Professor Qiang Li and Fang Zuo for their support and discussions. G.C. expresses his warmest thanks to the Chinese Academy of Sciences for the granting of a Visiting Professorship for Senior International Scientists at NAOC. This work is supported by the National Natural Science Foundation of China (grant No: 61273248, 61075033), the Natural Science Foundation of Guangdong Province (2014A030313425, S2011010003348), and the Open Project Program of the National Laboratory of Pattern Recognition (NLPR, 201001060).

\end{document}